%% file: agnlp_co.tex
\documentclass[usenatbib,useAMS]{mn2e}
\usepackage{graphicx}
\usepackage{aas_macros}
\usepackage{amssymb}
\newcommand{\kms}{km s$^{-1}$}
\newcommand{\ergs}{erg s$^{-1}$}
\newcommand{\mics}{$\mu$m}
\newcommand{\msun}{M$_{\odot}$}

\newcommand{\lx}{L$_{\rm X}$}
\newcommand{\smass}{M$_{\star}$}
\newcommand{\ico}{I$_{\rm CO}$}
\newcommand{\icor}{I$_{\rm CO, 25}$}
\newcommand{\lco}{L$^{\prime}_{\rm CO}$}
\newcommand{\lgal}{$L_{\rm GAL}$}
\newcommand{\lagn}{$L_{\rm AGN}$}
\newcommand{\alphaco}{$\alpha_{\rm CO}$}
\newcommand{\taud}{$t_{\rm dep}$}
\newcommand{\coline}{CO 2$\rightarrow$1}

\title[CO in LLAMA galaxies]{LLAMA: Normal star formation efficiencies of molecular gas in the centres of luminous Seyfert galaxies}

\author[D. J. Rosario et al.]
{D.J.~Rosario$^{1}$\thanks{Email: david.rosario@durham.ac.uk},
L.~Burtscher$^{2,3}$,
R.I.~Davies$^{2}$,
M.~Koss$^{4}$,
C.~Ricci$^{5,6,7}$, 
D.~Lutz$^{2}$,
R.~Riffel$^{8}$,
\newauthor 
D.M.~Alexander$^{1}$,
R.~Genzel$^{2}$,
E.H.~Hicks$^{9}$,
M.-Y.~Lin$^{2}$ ,
W.~Maciejewski$^{10}$,
\newauthor 
F.~M\"uller- S\'anchez$^{11}$,
G.~Orban de Xivry$^{12}$,
R.A.~Riffel$^{13}$,
M. Schartmann$^{14,15,2}$,
\newauthor 
K. Schawinski$^{16}$,
A. Schnorr-M\"uller$^{8}$,
A.~Saintonge$^{17}$,
T.~Shimizu$^{2}$,
A.~Sternberg$^{18}$,
\newauthor 
T.~Storchi-Bergmann$^{8}$,
E.~Sturm$^{2}$,
L.~Tacconi$^{2}$, 
E. Treister$^{5}$,
and S.~Veilleux$^{19}$
\\
$^1$Department of Physics, Durham University, South Road, DH1 3LE, Durham, UK \\
$^2$Max-Planck-Institut f\"ur extraterrestrische Physik, Postfach 1312, D-85741, Garching, Germany \\
$^3$Sterrewacht Leiden, Universiteit Leiden, Niels-Bohr-Weg 2, 2300 CA Leiden, The Netherlands \\
$^4$Eureka Scientific, Inc., 2452 Delmer Street Suite 100, Oakland, CA 94602-3017 \\
$^5$Instituto de Astrofisica, Pontificia Universidad Cat\'olica de Chile,Vicu\~na Mackenna 4860, Santiago, Chile \\
$^6$Chinese Academy of Sciences South America Center for Astronomy and China-Chile Joint Center for Astronomy,  \\
 $\phantom{6}$ Camino El Observatorio 1515, Camino El Observatorio 1515, Las Condes, Santiago, Chile \\
$^7$Kavli Institute for Astronomy and Astrophysics, Peking University, Beijing 100871, China \\
$^8$Departamento de Astronomia, Universidade Federal do Rio Grande do Sul, IF, CP 15051, 91501-970 Porto Alegre, RS, Brazil \\
$^9$Department of Physics \& Astronomy, University of Alaska Anchorage, AK 99508-4664, USA \\
$^{10}$Astrophysics Research Institute, Liverpool John Moores University, IC2 Liverpool Science Park, 146 Brownlow Hill, L3 5RF, UK \\
$^{11}$Center for Astrophysics and Space Astronomy, University of Colorado, Boulder, CO 80309-0389, USA \\
$^{12}$Space Sciences, Technologies and Astrophysics Research Institute, Universit\'e de Li\`ege, All\'ee du Six Ao\^ut 19C, 4000 Li\`ege, Belgium\\
$^{13}$Departamento de F\'isica, Centro de Ci\~encias Naturais e Exatas, Universidade Federal de Santa Maria, 97105-900 Santa Maria, RS, Brazil \\
$^{14}$Centre for Astrophysics and Supercomputing, Swinburne University of Technology, Hawthorn, Victoria, 3122, Australia \\
$^{15}$Universit\"ats-Sternwarte M\"unchen, Scheinerstrasse 1, D-81679 M\"unchen, Germany \\
$^{16}$Institute for Astronomy, Department of Physics, ETH Zurich, Wolfgang-Pauli-Strasse 27, CH-8093 Z\"urich, Switzerland \\
$^{17}$Department of Physics and Astronomy, University College London, Gower Street, London, WC1E 6BT, UK\\
$^{18}$Raymond and Beverly Sackler School of Physics \& Astronomy, Tel Aviv University, Ramat Aviv 69978, Israel \\
$^{19}$Department of Astronomy and Joint Space-Science Institute, University of Maryland, College Park, MD 20742-2421 USA
}

\begin{document}

\maketitle

\begin{abstract}

Using new APEX and JCMT spectroscopy of the \coline\ line, we undertake a controlled study of cold molecular
gas in moderately luminous (L$_{bol} = 10^{43-44.5}$ \ergs) Active Galactic Nuclei (AGN) 
and inactive galaxies from the Luminous Local AGN with Matched
Analogs (LLAMA) survey. We use spatially resolved infrared photometry of the LLAMA galaxies
from 2MASS, {\it WISE}, {\it IRAS} \& {\it Herschel}, corrected for nuclear emission using multi-component spectral energy distribution (SED) fits, 
to examine the dust-reprocessed star-formation rates (SFRs), molecular gas fractions and star formation
efficiencies (SFEs) over their central 1--3 kpc. We find that the gas fractions and central SFEs of both active and inactive galaxies are similar
when controlling for host stellar mass and morphology (Hubble type). The equivalent central molecular gas depletion times are consistent 
with the disks of normal spiral galaxies in the local Universe. Despite energetic arguments that the AGN in LLAMA should be capable
of disrupting the observable cold molecular gas in their central environments, our results indicate that nuclear radiation only couples weakly with this phase. 
We find a mild preference for obscured AGN to contain higher amounts of central molecular gas,
which suggests a connection between AGN obscuration and the gaseous environment of the nucleus. 
Systems with depressed SFEs are not found among the LLAMA AGN. We speculate that the processes that sustain 
the collapse of molecular gas into dense pre-stellar cores may also be a prerequisite for the inflow of material on to AGN accretion disks. 

\end{abstract}

\section{Introduction}

Active Galactic Nuclei (AGN\footnote{We use the acronym `AGN' for both singular and plural forms.})  
are important phases in the life cycle of a galaxy, during which its
central supermassive black hole (SMBH) accretes material from the circum-nuclear environment (the inner few 100s of pc).
The amount of material needed to fuel an AGN is not large; a powerful Seyfert galaxy with a
fiducial bolometric luminosity of $10^{45}$ \ergs\ can be sustained for a Myr through the accretion of only  
$\approx 2\times10^{5}$ \msun\ of gas (assuming a characteristic radiative efficiency of accretion $\eta_{r} = 10$ per cent). 
This constitutes a minute fraction of the gas usually available in the central regions of spiral galaxies, but 
as it falls into the deep potential well of an SMBH, it can gain a tremendous amount of energy,
some of which will be liberated in the form of radiation, winds, and relativistic jets. The coupling
of the liberated energy (or momentum) with extended gas is believed to significantly impact the host galaxy,
both in the circum-nuclear environment, where it is responsible for the regulation of SMBH scaling relationships, 
and if the AGN is powerful enough, over the entire host (see \citet{fabian12} and references therein). Indeed,
AGN feedback is the crucial input needed in galaxy formation theory to explain the shut-down of star formation in massive galaxies
over the history of the Universe \citep{bower06, croton06,somerville08}.

Cold molecular gas is the primary raw material for the formation of new stars (see \citet{kennicutt12} for a recent extensive review).
The incidence and surface density of molecular gas is closely related to the star-formation rate (SFR) within galaxies \citep{bigiel08, schruba11}.
Molecular gas scaling laws, as these relationships have been widely denoted, exist even over regions as small as $\approx1$ kpc 
in star-forming galaxies \citep{leroy13}, and their exact form is known to depend on the compactness and intensity of the star-formation
associated with the gas \citep[e.g.][and references therein]{genzel10}. The efficiency with which stars form in molecular clouds (the
`star formation efficiency'; hereafter SFE) is moderated by their turbulent support against self-gravitational collapse \citep[e.g.][]{krumholz05}, 
which is influenced by a number of physical and dynamical processes. In this work, we examine whether energy from an AGN
has an important impact on the SFE in the central regions of galaxies. This is an important test of the ability
of AGN to directly influence the material responsible for star formation in galaxies. 

Our study considers the nature of molecular gas within 1--3 kpc of the nucleus\footnote{We will use the term `central' consistently to refer
to the inner few kpc of a galaxy.} in moderately luminous nearby Seyfert galaxies 
(Section \ref{co_data}). This physical scale is large enough to average over many molecular clouds and is therefore 
insensitive to stochasticity in the SFE due to variations in the dense molecular component between individual clouds \citep{lada12}. 
However, we are also looking at a region of these galaxies small enough that the energetic feedback from the current phase
of accretion in their AGN can potentially drive out or influence a major part of the cold gas (Section \ref{feedback_calcs}).

The global SFE is known to be a function of gross galaxy properties \citep{saintonge11, saintonge12},
such as stellar mass, level of disturbance and offset from the so-called `Galaxy Main Sequence of star formation' 
\citep[e.g.][]{speagle14}. It is well-established that AGN host galaxies have distinguishing properties: they tend to be
massive galaxies of intermediate Hubble type \citep[large disks and bulges; e.g.][]{adams77,schawinski10}, 
and show higher rates of dusty star-formation than other galaxies of similar mass \citep[e.g.][]{dahari88,kauffman03,rosario16}. 
Therefore, any investigation into the direct influence of AGN on the nature of molecular gas should be sensitive
to the particularities of AGN hosts. As a trivial example, local AGN and inactive galaxies selected solely on the basis of
a magnitude-limited catalog, such as the RC3 \citep[Third Reference Catalogue of Bright Galaxies;][]{devaucouleurs91}, 
will differ in their stellar mass distributions. Since the SFE decreases with stellar mass \citep{saintonge11}, the median
SFE among the AGN of this hypothetical sample would be lower than that of the inactive galaxies, purely due to the bias
of the AGN host population. To overcome this, our study has adopted a careful control strategy for overall galaxy properties,
to help discern the possible effect of AGN feedback on central molecular gas independent of systematic AGN-independent trends (Section \ref{sample_description}).

Cold molecular gas in galaxies reveals itself most effectively through emission in the rotational lines of the polar diatomic
$^{12}$CO molecule, an abundant component of metal-enriched interstellar molecular gas \citep[e.g.,][]{solomon75}.
Since the advent of millimetre-wave molecular spectroscopy in astronomy, the low order $^{12}$CO rotational lines have been the principal
tracer of the bulk of the molecular gas in galaxies \citep[e.g.][]{young91}. 
The use of these features, however, comes with a degree of complexity. In most
circumstances, the low order CO lines are optically thick, so their use as a mass proxy, through the CO-to-H$_{2}$ conversion 
factor \alphaco, is sensitive to the metallicity, geometry, and cloud structure of the molecular gas emitting the lines \citep{genzel12,bolatto13}. In general,
these properties are not known and have to be assumed. A common approach is to adopt an average \alphaco\ found for the disk
of the Milky Way, though there is considerable real variation in this factor in the disks of galaxies \citep{sandstrom13}. In addition, the
centres of galaxies often show depressed \alphaco, which, if not taken into account, could mistakenly imply incorrectly
low gas masses, and by extension, incorrectly high SFEs in these environments. In this study, we use a refined statistical approach
that propagates our uncertain knowledge of essential conversions such as \alphaco\ to quantify the differences in gas
fractions (Section \ref{gas_fracs}) and SFEs (Section \ref{sf_efficiency}) in the centres of luminous Seyferts and inactive galaxies.

Local AGN have been the target of several CO studies, 
mostly with single-dish telescopes with beams that covered most of the emission from the host galaxies
 \citep{heckman89, maiolino97, curran01, bertram07}. 
 In recent years, interferometry is increasingly being used for small-scale studies of 
kinematics and outflows in Seyfert galaxies \citep[e.g.][]{garciaburillo05,garciaburillo14}, a field set to burgeon with
the advent of ALMA.

Early work reported differences in the gas content and SFE 
 between Seyfert galaxies and inactive galaxies, as well as differences across the various types of Seyferts \citep{heckman89}. Since then, 
other large sample studies have suggested that these results were driven by selection effects and that Seyfert hosts
actually have normal SFEs \citep[e.g.][]{maiolino97}. On the other hand, among more luminous local AGN, \citet{bertram07} 
reported enhanced SFEs intermediate between normal galaxies and strong starbursts, suggesting a connection between
the phenomenon of luminous AGN fuelling and the ultra-luminous dusty star-forming galaxies. 
A major complexity when interpreting these papers is that the characteristic host properties of the AGN were not always
adequately controlled for when making comparisons of inactive galaxies. 
There is also uncertainty in the level to which the AGN contaminates some of the FIR emission 
that was used to estimate the SFR, particularly in studies of luminous systems. This work adopts a careful
control strategy and a uniform multi-wavelength analyses to jointly compare the SFE of AGN and inactive galaxies.

Section \ref{data} describes our experimental setup, including sample selection and the compilation of essential measurements.
Section \ref{stats} outlines our statistical approach, followed by the examination of key results in Section \ref{results}. We
discuss the interpretation of our findings in Section \ref{discussion}. Throughout this paper, we assume a WMAP9 concordance
cosmology \citep{hinshaw13} as adapted for the Astropy\footnote{http://www.astropy.org/} Python package \citep{astropy}. Unless otherwise
specified, a Chabrier IMF \citep{chabrier03} is assumed for stellar population-dependent quantities. All quoted uncertainties
are equivalent to 1 standard deviation, and we adopt a threshold probability of 5 per cent when evaluating the statistical significance
of a difference from a Null Hypothesis.

\section{Data and Measurements} \label{data}

The properties of our sample of AGN and inactive galaxies are described below, followed by a description of new CO spectroscopy,
archival CO data, multi-wavelength photometry and the spectral energy distribution (SED) fitting method used to derive IR
luminosities of the dust heated by star-formation and the AGN. Most of datasets used in this work can
be obtained from public repositories. Raw and reduced APEX CO spectral data may be obtained directly from the lead author. 

\subsection{The LLAMA sample} \label{sample_description}

The Luminous Local AGN with Matched Analogs (LLAMA\footnote{http://www.mpe.mpg.de/llama}) project has targetted 20 southern 
AGN and a set of 19 inactive galaxies that serve as a carefully-selected control sample, to explore the relationships
between on-going nuclear activity and circum-nuclear dynamics and star-formation. The continuing program will  
obtain {\it HST} imaging, high-spatial resolution AO-assisted near-infrared (NIR) integral-field unit spectra with the VLT/SINFONI
and high S/N contiguous optical-NIR IFU spectra with VLT/XSHOOTER for all sources. The AGN were selected
on the basis of their luminosities in the 14--195 keV band, from the {\it SWIFT}-BAT all-sky survey.  
The choice of ultra-hard X-ray selection ensures a minimal sensitivity to modest levels of X-ray obscuration, with
obscuration-dependent incompleteness only becoming important at equivalent Hydrogen column densities $N_{H} > 10^{24}$ cm$^{-2}$, i.e.,
in the Compton-thick regime \citep{ricci15}. The AGN in LLAMA are comprised of a volume-limited sample with 
$z<0.01$, L$_{BAT} > 10^{42.5}$ \ergs, and declination $\delta < 15^{\circ}$. The inactive control sample are 
galaxies with no known signatures of nuclear activity, selected from the RC3 \citep{devaucouleurs91} 
to satisfy the same observability criteria and redshift limit as the AGN, and are matched to them 
within $\pm0.2$ dex in H-band luminosity, $\pm1$ in RC3 Hubble type, and $\pm 15^{\circ}$ in galaxy inclination. 
They tend to be at somewhat lower distances than the AGN \citep[][also see Figure \ref{lco_distance} of this paper]{davies15}.

\begin{figure*}
\includegraphics[width=\textwidth]{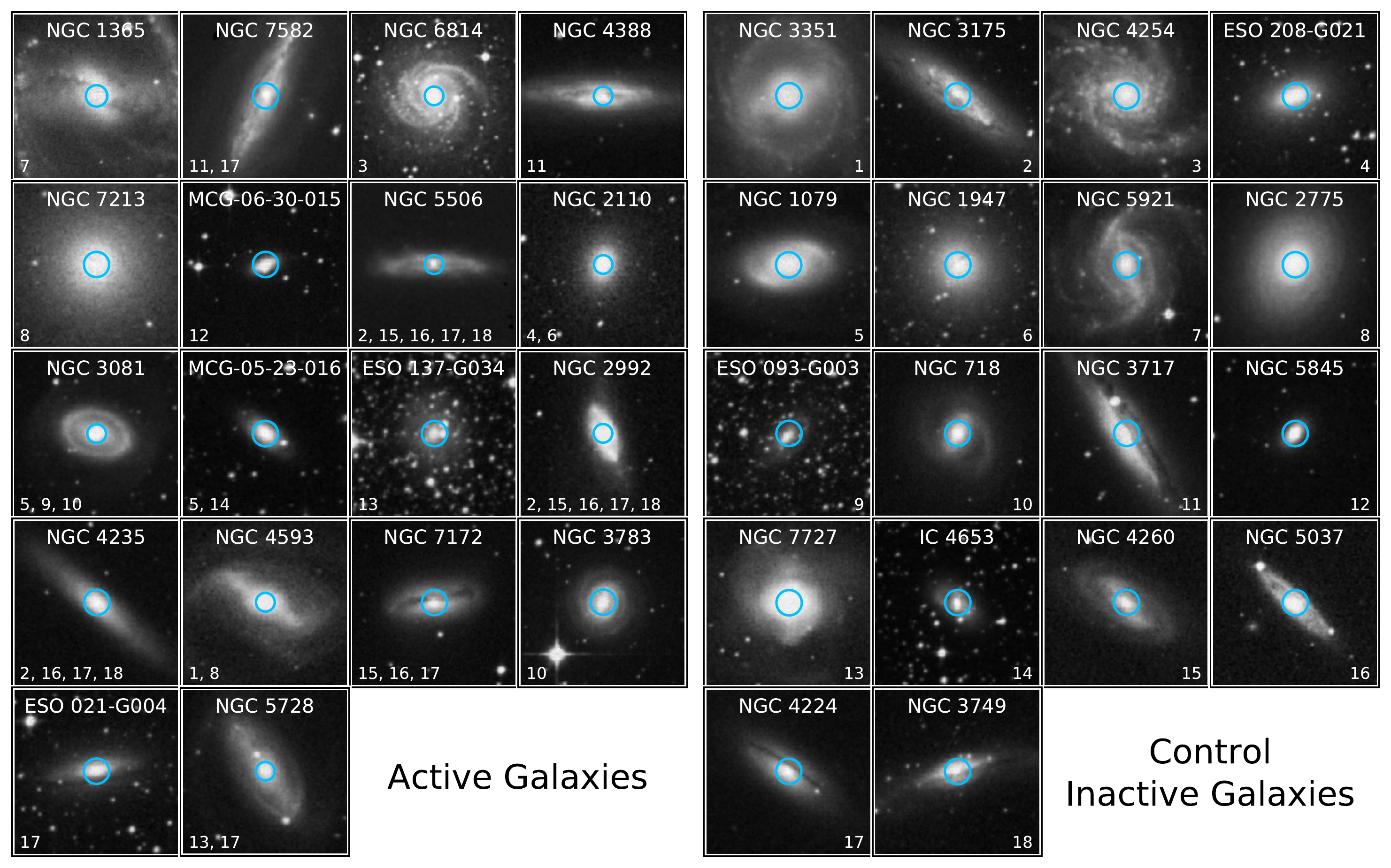}
\caption{ A gallery of DSS-R images of the galaxies from the LLAMA survey featured in this study. Each panel is $3\arcmin \times 3\arcmin$ in size.
The half-power circular beam of the single-dish radio telescope used to obtain CO data for each galaxy is shown as a cyan circle placed on
the pointing centre of the observation.  AGN hosts are shown on the left and control galaxies on the right, with both sets ordered by distance 
increasing across the panels left to right and down.
The numbers in the lower right of each panel of the control galaxy images serves as an index:
each AGN is matched to one or more inactive galaxies, with their corresponding
indices shown in the lower left of each of the AGN host images.
}
\label{dss_montage}
\end{figure*}

Figure \ref{dss_montage} displays DSS R-band images of the 36 LLAMA galaxies used in this study. 
Centaurus A (NGC 5128) was excluded from the study due to its proximity -- it is a third of the distance
of the next nearest LLAMA AGN. Its unique matched analog, NGC 1315, was also removed. 
MCG-05-14-012 (ESO 424-12) was excluded because its low stellar mass and large distance implied
unreasonably long observing times to satisfy our CO characterisation criteria (see below).

The AGN and control galaxies in Figure \ref{dss_montage} are ordered by distance. A number key system is
used to indicate the AGN that are matched to each control galaxy. A single control galaxy may be matched to
more than one AGN within our matching tolerances, and reciprocally, an AGN may have more than one control galaxy. A visual examination
of Figure \ref{dss_montage} is a useful exercise to ascertain the scope and accuracy of our matching approach.

The AGN in LLAMA are nearby well-studied Seyfert galaxies with a wealth of contextual data. Spectral classifications
from the literature were compiled in \citet{davies15} and in Table \ref{basic_data}.
The classifications cover the traditional Seyfert 1-1.8 and Seyfert 2 categories, but also include Seyfert 1i, which
only show broad permitted lines in the infrared, and Seyfert 1h, which show broad optical lines in polarised light.
In addition to optical classifications,  we obtained absorption-corrected rest-frame 2--10 keV fluxes for our AGN from 
\citet{ricci17}, and converted these to intrinsic  2--10 keV X-ray luminosities (\lx), 
using the compilation of redshift-independent distances from \citet{davies15}. \citet{ricci17} also provides 
estimates of the intrinsic line-of-sight absorbing column densities (N$_{\rm H}$) towards the nucleus for all the AGN.
In objects with no hint of intrinsic X-ray absorption, N$_{\rm H}$ is set to an upper limit of $10^{20}$ cm$^{-2}$.

\subsection{\coline\ spectroscopy} \label{co_data}

\begin{figure*}
\includegraphics[width=\textwidth]{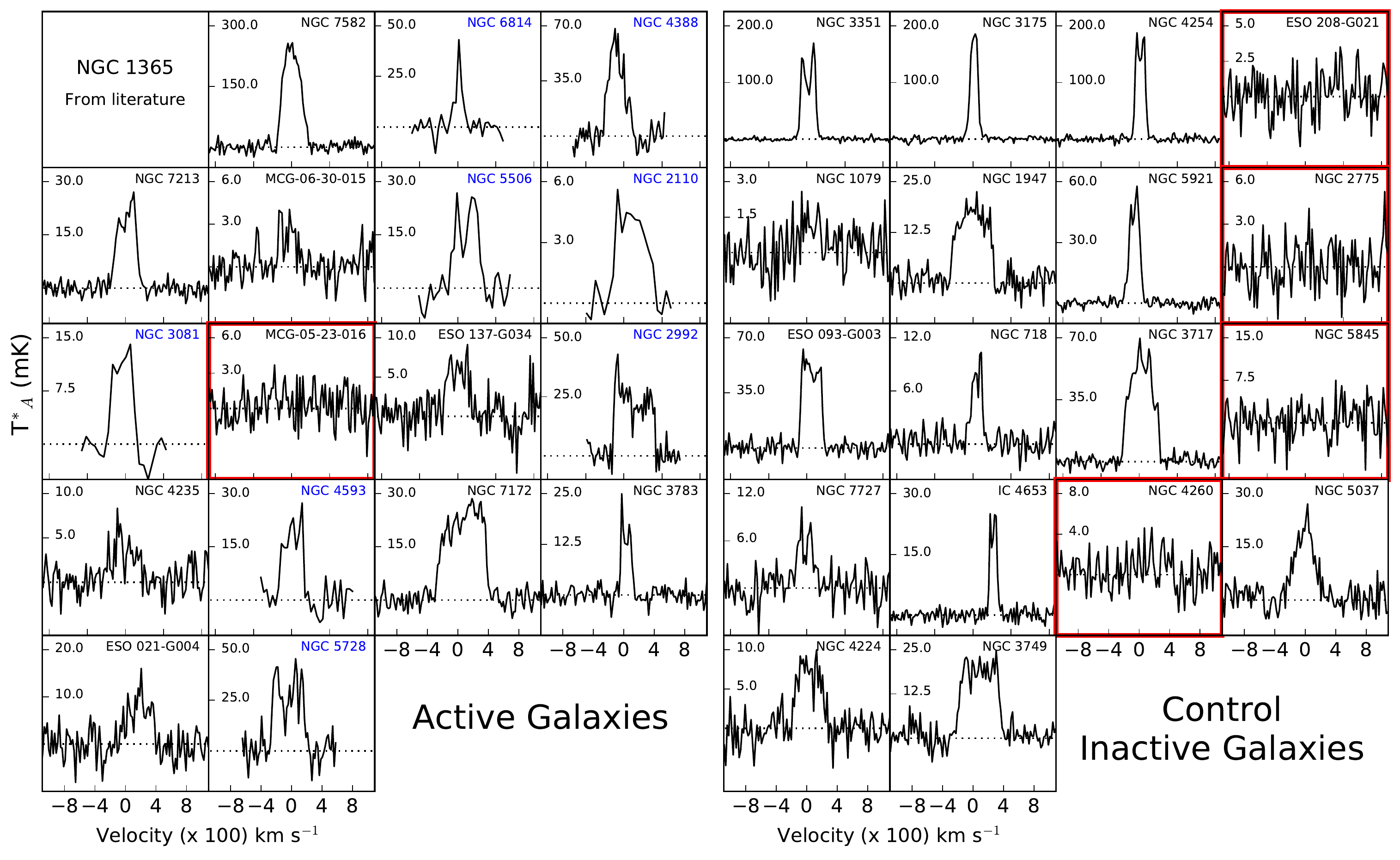}
\caption{ \coline\ line spectra of the galaxies from the LLAMA survey featured in this study. Each panel spans a fixed range of 2200 \kms\ of velocity
around the systemic velocity of the galaxy to allow a simple visual comparison of the kinematics of the lines. Fluxes are expressed as
atmosphere-corrected antenna temperatures (T$^{\star}_{\rm A}$). 
AGN hosts are shown on the left and control galaxies on the right, following the same order as Figure \ref{dss_montage}. 
Black and blue labels for the names respectively indicate the galaxies for which APEX and JCMT observations were taken. 
We relied on literature measurements of \coline\ for NGC 1365 and do not plot a spectrum here.
Panels with red boundaries mark the targets which are not detected in CO at SNR$<3$.
}
\label{co_montage}
\end{figure*}

For this study, we observed or compiled archival spectroscopy of the $^{12}$\coline\ line at a rest frequency of $230.538$ GHz,
which achieves an optimal trade-off between spatial resolution and line sensitivity. Our targets
are nearby, so we used 12-m class single-dish millimetre telescopes for our observations. The approximate half-power beam widths (HPBWs)
achieved for each target are shown as cyan circles in Figure \ref{dss_montage}. They span the central regions of our targets, subtending
projected radii of 0.7-2.7 kpc over the full range of distances of the sample. Figure \ref{co_montage} presents a montage of the final reduced CO spectra
of the LLAMA targets.

\subsubsection{APEX spectroscopy}

27 targets were observed in a dedicated LLAMA follow-up survey with the Atacama Pathfinder Experiment (APEX) telescope 
(Program ID M0014\_96; PI: Rosario). Spectra were taken in several tracks over October -- December 2015, with varying but generally favourable
conditions. The requested final integration time of a target was designed to achieve either a S/N $\ge5$ detection of the \coline\
line, based on its central FIR luminosity and a typical star-formation depletion time of 1 Gyr (see Section \ref{sf_efficiency} for definitions), 
or a limit on the central molecular gas fraction of 5 per cent. This approach is preferred over a fixed depth survey, since it allows us to devote
more observing time to gas-poor and distant systems, and achieve a volume-limited equivalent CO survey that mirrors the LLAMA selection strategy. 

The APEX-1 heterodyne receiver system was used with a standard beam switching sequence. The equivalent circular
half-power beam width (HPBW) of this setup is 27\farcs1 at 230 GHz. Calibrators were chosen 
following standard APEX queue observing guidelines. 
All the spectra were reduced with the CLASS software from the GILDAS package\footnote{https://www.iram.fr/IRAMFR/GILDAS/}.
Spectra from contiguous tracks on the same observation date were resampled and added into a single spectrum to which 
baseline corrections were applied. When the \coline\ line was detectable in spectra from different dates, we visually compared
them to ascertain if any flux calibration or wavelength calibration systematics were evident. Finding none, we proceeded to
combined baseline-subtracted spectra taken on different dates into a single final spectrum for each object.

The spectra are shown in Figure \ref{co_montage}.
We adopted a two-tier approach to measure the integrated CO flux from a spectrum. We first fit\footnote{We rely on the versatile Python LMFIT package 
with a least-squares algorithm for line profile fits.} the CO line with a single gaussian profile,  
which allowed a preliminary assessment of the line strength, centre and width. In galaxies with well-detected lines, 
several cases of substantial deviations from a simple gaussian profile are evident, implying that a single gaussian fit will not capture
the line flux accurately. Therefore, for lines with a preliminary SNR $>15$, we remeasured the line flux as follows.
We integrated the spectrum within $\pm10\times \sigma_{est}$ of the CO line centre to obtain the total flux, where both
the line width $\sigma_{est}$ and its centre come from the gaussian fit. 
We estimated the spectral noise from the baseline variance in spectral regions with absolute velocity offsets
$> 1000$ \kms\ from the line centre. We were able to adopt this technique of integrated flux measurements
for strongly detected lines due to the very flat baselines and wide bandpass of our APEX spectra.

We converted antenna temperatures to luminance units (Jy \kms) using a fixed conversion of 39 Jy/K, suitable for APEX at 230 GHz.

\subsubsection{JCMT data}
8 AGN were observed with the James Clerk Maxwell Telescope (JCMT) in a filler program
between February 2011 and April 2013. The A3 (211-279 GHz) receiver was used with a beam size of 20\farcs4.  
Each galaxy was initially observed for 30 minutes. For weak detections, additional observations were obtained up to no more than 2 hours. 
The individual scans for a single galaxy were first-order baseline-subtracted 
and then co-added. The short bandpass of the JCMT spectra do not include enough line-free regions for an 
estimation of the spectral noise. Therefore, we fit the line spectra with a combination 
of two gaussian profiles and a flat continuum, and add a 10 per cent error in quadrature to the uncertainty on the line fluxes to account for any
spectral baselining uncertainties. We used a conversion factor of 28 Jy/K (an aperture efficiency of 0.55) to scale from antenna temperatures
to luminance units.

\subsubsection{Measurements from the literature}
NGC 1365, a Seyfert 1.8 in a massive barred spiral galaxy, has been the subject of extensive CO follow-up in the literature.
To match the approximate depths and resolutions of the CO data for the rest of the LLAMA sample, 
we adopt the \coline\ flux measurement of NGC 1365 from \citet{curran01}, based on spectroscopy with the 15-m Swedish 
ESO Sub-millimetre Telescope (SEST; HPBW $=23$\arcsec\ at 230 GHz). They report a line flux of $6150\pm 410$ Jy \kms.

\input{Table1.tex}

\subsection{Infrared photometry}

Essential insight into the conditions of the molecular gas in our galaxies comes from a comparison of the cold gas masses with stellar masses
and the SFR in their central regions. We relied on infrared (IR) photometry and multi-component fits to the IR SED to estimate these quantities.
The near-IR in our galaxies is dominated by stellar light, while the mid-IR and far-IR/sub-millimeter bands give us a good handle on the
dust emission from the AGN `torus' and from star-forming regions. 

\subsubsection{Near-infrared photometry}

\begin{figure}
\includegraphics[width=\columnwidth]{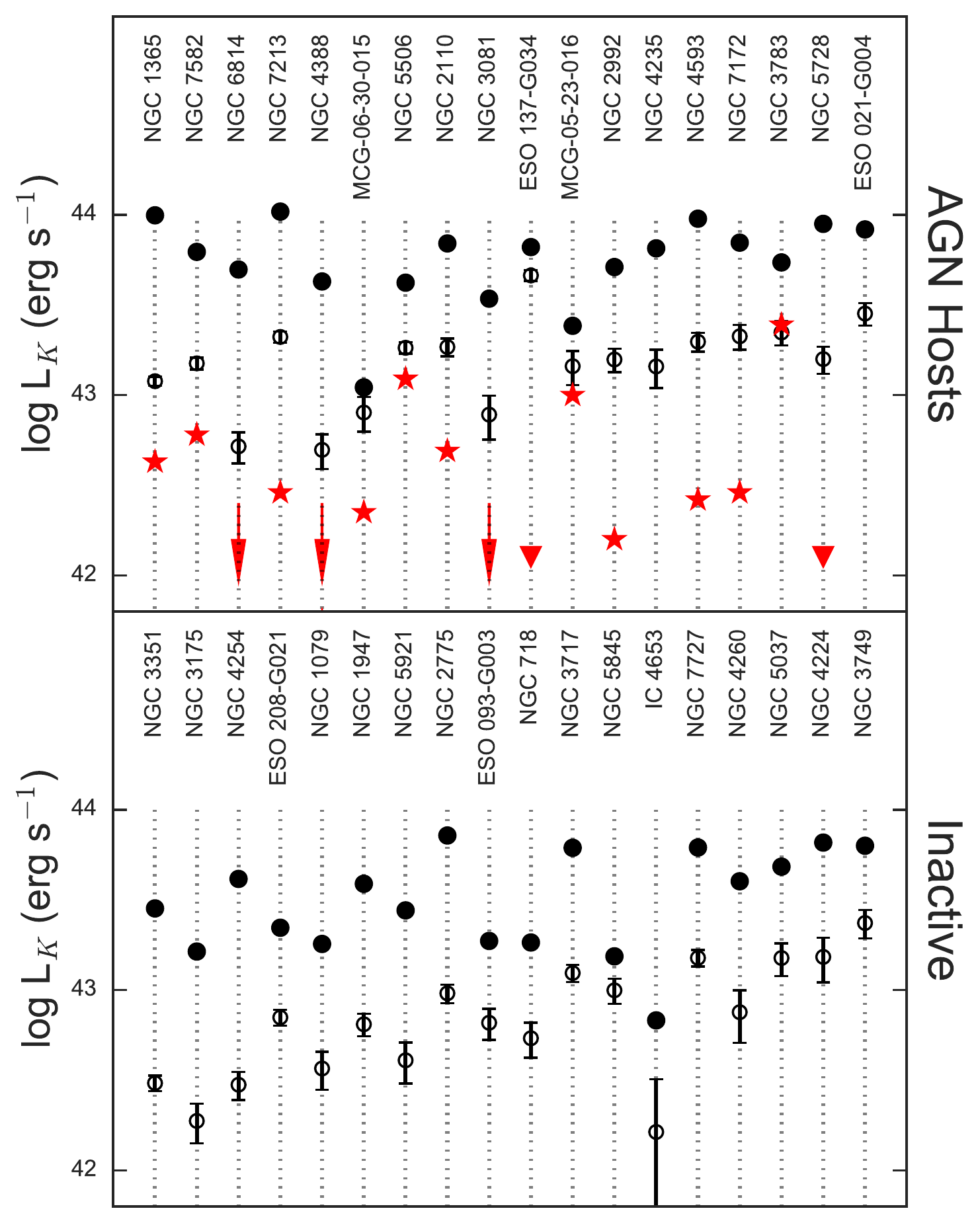}
\caption{K-band luminosities ($L_{K}$) of LLAMA AGN (top set) and inactive galaxies (bottom set). Both sets are individually ordered by increasing
distance from left to right. Integrated luminosities (full circle symbols) and luminosities scaled to the CO telescope beam (open circle symbols)
are from 2MASS data. Estimates of the pure AGN luminosity in the K-band from \citet{burtscher15} are shown as red points (stars for measurements,
small arrowheads for upper limits). The AGN luminosity measurements for NGC 6814, NGC 4388 \& NGC 3081 lie outside the plotted range, and this 
is indicated by the large red arrows. 
}
\label{Kband_lums}
\end{figure}

Images and photometry in the near-infrared (NIR) J/H/K$_{s}$ bands for all our galaxies were compiled 
from the 2MASS survey through the NASA/IPAC Infrared
Science Archive (IRSA)\footnote{http://irsa.ipac.caltech.edu/Missions/2mass.html}. For total galaxy photometry, we used measurements
from the 2MASS Extended Source Catalog (XSC) and the Large Galaxy Atlas, both of which employ light profile fits to the galaxies
to yield integrated fluxes that are robust to the presence of foreground stars, or variations in the seeing or background. 

We performed beam-matched photometry directly from the 2MASS K$_{s}$ images to obtain a measure of the near-IR light from our 
galaxies co-spatial with the molecular gas from our single-dish spectra. We assumed a circular gaussian beam
with a width ($\sigma_{\rm b}$) equal to that of the beam of the respective telescopes at 230 GHz. We
weighted the 2MASS K$_{s}$ by this beam profile, and integrated over
the images, scaling by the appropriate 2MASS zeropoints, to derive a K$_{s}$ flux matched to the single dish beam. This approach is 
appropriate because the equivalent resolution of the 2MASS images is a few arcseconds, much smaller than the beam widths of the
single-dish observations. Figure \ref{Kband_lums} displays the integrated and beam-matched luminosities in the K$_{s}$
for all the LLAMA galaxies, indicating that the single-dish beams sample a few tens of percent of the total NIR emission in these systems.

\citet{burtscher15} characterised the dilution of the stellar photospheric CO bandhead (2.3 \mics) in a number of active galaxies, 
enabling a very sensitive measurement of the NIR luminosity of their AGN. This work has published constraints for
15 of the 18 LLAMA AGN and we adopted these measurements of the intrinsic K band luminosity of their nuclear sources.
For the remaining 3 AGN,  we adopted the AGN's K-band luminosity estimated from our multi-component SED fits (Section \ref{sed_fits}).
ESO 137-G034 \& NGC 5728, two heavily obscured AGN, do not appear to show any AGN light in the K-band to the limit of the \citet{burtscher15}
analysis, probably due to a high optical depth to NIR radiation in their AGN tori. Our estimates/limits on the AGN's NIR luminosities
are plotted in Figure \ref{Kband_lums} as red star/arrowhead points. In most cases, they are a few times weaker than the beam-matched K-band luminosities,
but in NGC 5506, MCG-05-23-016 \& NGC 3783, the AGN dominates the central emission.

We subtracted the contribution of the AGN light from the beam-matched 2MASS K$_{s}$ fluxes, assuming
a flat AGN SED across the K-band. The resulting pure stellar K-band luminosity is used in the study of the central
gas fraction (Section \ref{gas_fracs}). If the AGN's luminosity was within -2$\sigma$ of the beam-matched central flux, we considered this flux
to be an upper limit on the central stellar luminosity.

\subsubsection{Mid-infrared photometry} \label{mir_phot}

\begin{figure}
\includegraphics[width=\columnwidth]{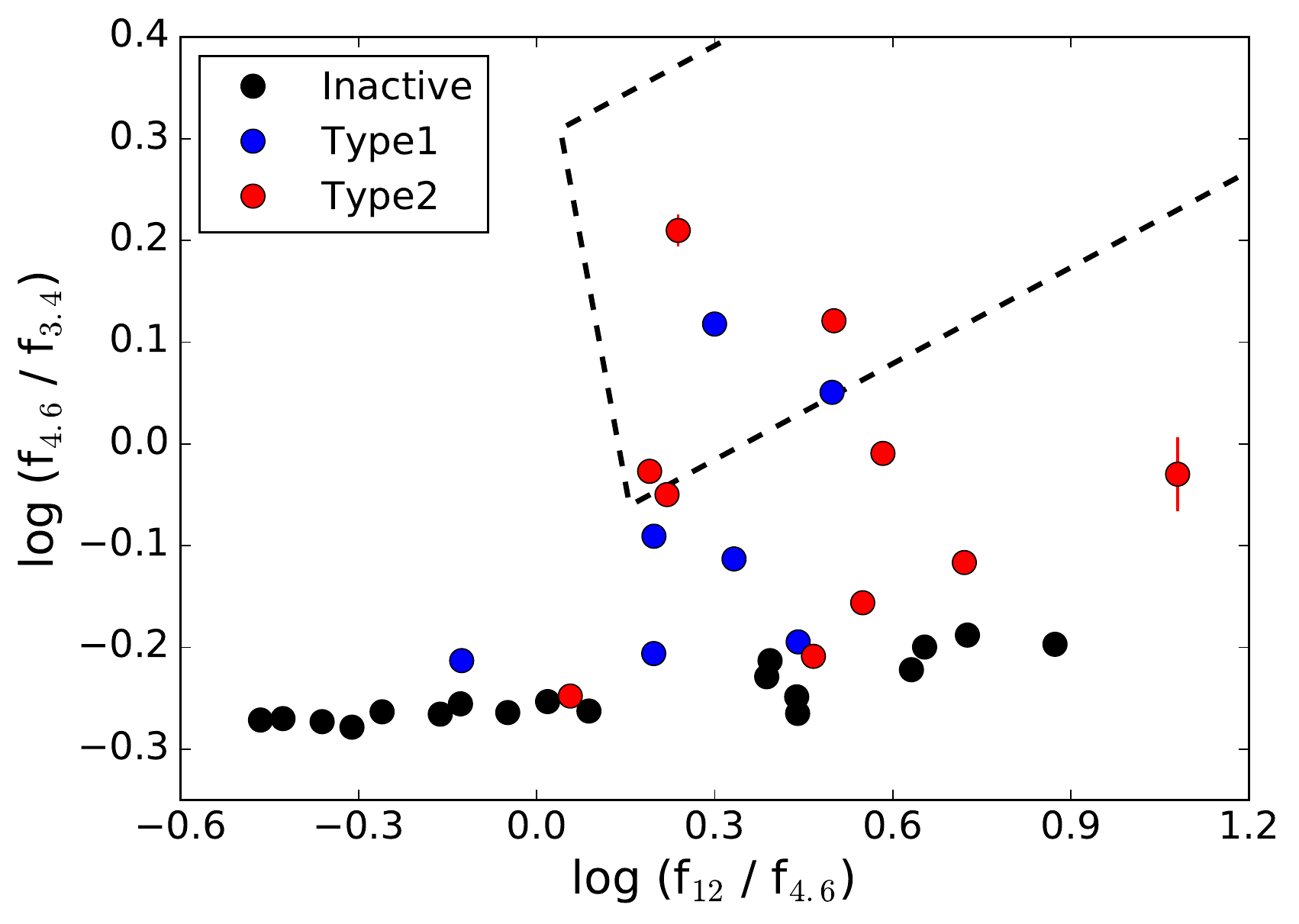}
\caption{ LLAMA galaxies plotted on the {\it WISE} colour-colour diagram of \citet{mateos12}, with the ratio of W3 (12\mics) to W2 (4.6 \mics)
fluxes on the x-axis and W2 to W1 (3.4 \mics) fluxes of the y-axis. AGN are shown
with blue (Type 1) and red (Type 2) points, and inactive galaxies with black points. Error bars are plotted, but may be too small to be visible.
The dashed lines delineate a region of the diagram which contains objects with a MIR SED dominated by AGN torus emission.
}
\label{wise_color-color}
\end{figure}

We compiled integrated mid-infrared (MIR) fluxes for our targets from the ALLWISE catalog available on IRSA. Associations with the catalog
were made using a circular cone search with a tolerance of 5 arcsec, which yielded a single counterpart within 2.2 arcsec in all cases. 

We adopted the {\it WISE} pipeline-produced ``GMAG" aperture photometry which relies on scaled apertures derived from the profile of the galaxy 
from the 2MASS XSC. The aperture photometry is preferred over the standard profile-fit ``MPRO" photometry from the ALLWISE catalog,
since all the LLAMA galaxies are moderately to well-resolved in the {\it WISE} Atlas images.

In Figure \ref{wise_color-color}, we compare our AGN and inactive galaxies in a diagram of flux ratios between 
{\it WISE} bands. Objects which are dominated by AGN emission in the MIR have been shown to lie within the region delineated
by dashed lines in such a diagram \citep{mateos12}. Many of the LLAMA AGN lie within or close to the AGN-dominated region, 
consistent with our selection of relatively luminous Seyferts. On the other hand, all the inactive galaxies lie well away from the
region, showing that there is no sign of any hot dust emission in our control sample. The lack of even heavily-obscured nuclear sources
in the control galaxies confirms their inactive nature. 

While most of our AGN have {\it Herschel} far-infrared photometry (Section \ref{fir_phot}), only two of the control galaxies have been
the target of {\it Herschel} imaging programs. For a measure of the resolved thermal infrared emission for the rest,
we rely on the {\it WISE} W4 (22 \mics) Atlas images. The control galaxies do not contain detectable AGN, so we can confidently assume that
any of their thermal emission in the long-wavelength MIR  arises from dust heated by stars. The PSF of the {\it WISE} Atlas images
in the W4 band has a FWHM of  $\approx 11\farcs8$  and varies between objects and epochs; 
its large size in comparison to the single-dish beam demands the use of uncertain deconvolution techniques for accurate beam-matched
photometry. We instead rely on a simplified measure of the thermal dust emission within the single-dish beam based on the following procedure.
We match the {\it WISE} W4 images to the single-dish resolution by convolving them with a circular spatial gaussian kernel of width $\sigma_{k}$ given
by:
\begin{equation}
\sigma_{k} \; = \; \sqrt{\sigma_{\rm b}^{2} - \sigma_{ir}^{2}}
\end{equation}

\noindent where $\sigma_{\rm b}$ is the millimeter telescope beam and $\sigma_{ir}$ is the typical gaussian-equivalent 
width of the {\it WISE} W4 Atlas images ($\approx 11\farcs8$). We then extracted photometry from the beam-matched {\it WISE} images
within a circular aperture with a diameter equal to the HPBW of the single-dish telescope used for each object.

\subsubsection{Far-infrared photometry} \label{fir_phot}

\begin{figure*}
\includegraphics[width=\textwidth]{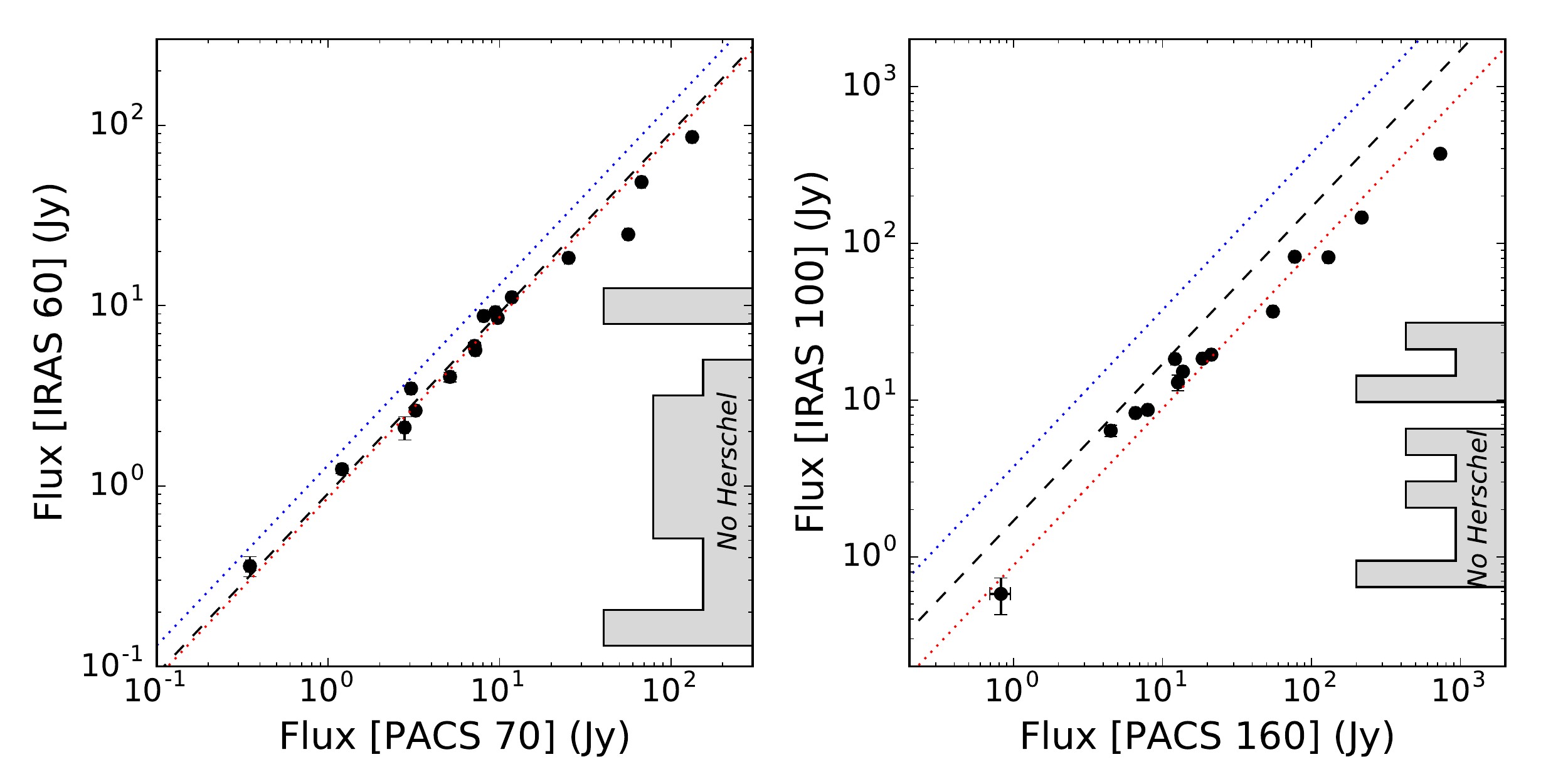}
\caption{A comparison of fluxes from {\it IRAS} and {\it Herschel}/PACS for sources in LLAMA with coverage from both facilities.
{\bf Left:} {\it IRAS} 60 \mics\ vs. PACS 70 \mics, {\bf Right:} {\it IRAS} 100 \mics\ vs. PACS 160 \mics. The lines in each panel
show the expected relationship for galaxy dust SED templates from \citet{dale02} spanning the full range of
the shape parameter $\alpha_{\rm D}  = 0.5 \textrm{--} 4.0$, with a typical value of $\alpha_{\rm D}  = 2.0$ shown
as the dashed line. The {\it IRAS} fluxes are compatible with the PACS fluxes, except at the bright end, where they tend
to be underestimated. The distribution of {\it IRAS} fluxes for the LLAMA sources that do not have {\it Herschel} coverage, including
most of the inactive galaxy subsample, are shown as vertical histograms in each panel.
}
\label{pacs_iras}
\end{figure*}

We employ both legacy {\it IRAS} data and modern {\it Herschel} data to get the best combined FIR datasets for the LLAMA galaxies. The AGN
have been the focus of targetted {\it Herschel} imaging programs, but most of the inactive galaxies do not have {\it Herschel} coverage. Therefore,
we have compiled all-sky {\it IRAS} data, where available, for the entire set, and compare these measurements to {\it Herschel} photometry
to understand and account for systematics between the datasets. Any large differences will complicate the controlled nature of our study, 
since {\it Herschel} maps, only widely available for our AGN, are more sensitive and allow better background subtraction than {\it IRAS} scans. 

Using the SCANPI facility from IRSA, we obtained 60 \mics\ and 100 \mics\ photometry from the Infrared Astronomical Satellite ({\it IRAS}) for 34
LLAMA galaxies. Two objects (MCG-05-23-016 \& NGC 3081) were flagged by IRSA to have problematic {\it IRAS} data. SCANPI is designed
for versatile use of {\it IRAS} all-sky data, allowing the user to choose between many different flavours of photometric measurements with
information about individual {\it IRAS} scans as well as combinations of all scans that cover a target.
We followed the recommendations in the SCANPI documentation for the working choice of photometric measurements. 
We only used data from median-combined scans. For sources with a flux $> 2$ Jy (SNR of several), 
we took the peak flux from the combined scans, except for sources which were determined to be extended, in which case, we used
a measure of the integrated flux (\textit{fnu\_t}). For weaker sources, we used photometry based on a point-source template fit, which
is more resilient to the complex background of the {\it IRAS} scans. In all cases, we adopted the estimate of the noise as the flux
uncertainty, and considered a source to be a detection if its flux was $>3\times$ the noise level. 

Sixteen AGN in our sample were observed with the {\it Herschel} Space Observatory \citep{pilbratt10} using the PACS
and SPIRE instruments, covering wavelengths from 70 \mics\ to 500 \mics. Details of these observations,
the reduction of the data and associated photometry are published in \citet[][PACS]{melendez14} and \citet[][SPIRE]{shimizu16a},
from which integrated photometric measurements and their uncertainties were obtained. 

Two inactive control galaxies (NGC 3351 \& NGC 4254) also have {\it Herschel} photometry from the KINGFISH survey of local galaxies
\citep{kennicutt11}. We adopted the KINGFISH photometric measurements for these galaxies from \citet{dale12}. 

In Figure \ref{pacs_iras}, we check for systematic zero-point differences between the {\it IRAS} and {\it Herschel}/PACS photometry 
by comparing photometry in nearby bands ({\it IRAS} 60 \mics\ vs. PACS 70 \mics, and {\it IRAS} 100 \mics\ vs. PACS 160 \mics). 
The various lines show the expected tracks of selected galaxy dust SED models from \citet{dale02}, 
and the vertical histograms show the distributions of {\it IRAS} fluxes for the remainder of the
sample that do not have {\it Herschel} coverage.

We find that the PACS fluxes of the brightest sources ($> 20$ Jy) are systematically brighter than their {\it IRAS} fluxes, 
both when compared to the trend shown by the fainter sources and against expectations from galaxy SED models. This is likely due 
to source emission that extends beyond the cross-scan width of the {\it IRAS} scans ($\approx 5'$) in some of the nearest and brightest
objects. The fainter sources lie in the range expected for typical cold dust SEDs of normal star-forming galaxies ($\alpha_{\rm D} = 2\textrm{--}4$; 
see Section \ref{sed_fits} for details). This suggests that any systematic offsets are minor. In addition, all LLAMA galaxies without
{\it Herschel} coverage, including most inactive galaxies, are $< 20$ Jy in both {\it IRAS} bands. Therefore, we can compare the FIR
properties derived for the AGN and inactive galaxies without major concerns about the disparity of their FIR data coverage.

In addition to the integrated photometry, we performed aperture photometry on the PACS 160 \mics\ images, when available, following a similar
procedure as described in Section \ref{mir_phot} for the {\it WISE} W4 images. The convolution kernel to match the PACS PSF to the single-dish beam
was calculated using an equivalent gaussian PSF with a FWHM of 11\farcs3  for the 160 \mics\ maps. 

\input{Table2.tex}

\section{Estimation and Statistics} \label{stats}

\subsection{Multi-component SED fitting} \label{sed_fits}

We fitted the infrared photometry of the AGN and galaxies using a multi-component Bayesian SED fitting package \citep[FortesFit;][]{rosario17}.
The fits combined three libraries of SED models: a) a set of single stellar population models (SSPs) generated
with the \citet{bruzual03} GALAXEV package; b) a single-parameter sequence of templates of the dust emission from galaxies 
heated by star-formation and the interstellar radiation field \citep{dale02}; c) a suite of empirical AGN template SEDs, covering
a range in MIR-to-FIR flux ratios.

The SEDs from the SSP models are parameterised by their age, chemical abundance and total stellar mass. In this study, we are only concerned with
the long-wavelength ($>1$ \mics) shape of the IR stellar emission from our galaxies, rather than the detailed properties of their stellar population.
This shape is only weakly affected by dust obscuration. Consequently, we did not consider any extinction when generating the library of SED models.
We considered a model grid of ten SSP ages logarithmically spaced between 5 Myr and 11 Gyr, and four metallicities 
with the solar metal abundance pattern but scaled to [Fe/H] of 1/50, 1/5, 1, and 2.5 of the solar value. The stellar mass is determined by the normalisation
of a particular SED model.

The galaxy dust emission templates of \citet{dale02} are a sequence of model SEDs for which a single parameter $\alpha_{\rm D}$, which is related
to the 60-to-100 \mics\ flux ratio of the template, has been shown to describe much of the variation observed in the MIR-to-FIR shape and the equivalent
width of PAHs among galaxies in the local Universe. For full flexibility, we allow the normalisation of the template, which determines 
\lgal, the integrated 8--1000 \mics\ luminosity of the galaxy's dust emission, to vary independently of $\alpha_{\rm D}$. 

For the AGN emission, we compiled a custom library of templates that span the range of mid-to-far IR empirical SED shapes reported
in the literature (Appendix \ref{agn_templates_discussion}). The library is parameterised by their integrated 8--1000 \mics\ luminosity arising from AGN-heated dust emission (\lagn) and
the ratio of the 160 \mics\ monochromatic luminosity to \lagn\ ($R_{160}$) which describes the steepness of the FIR tail of this emission.

Here we briefly summarise the key features of the fitting package, referring the interested reader to a forth-coming publication that fully documents
the software \citep{rosario17}.  

The parameters of the fit are treated as continuous variables and the routine is able to evaluate a hybrid SED model
(a combination of SEDs from all three components) at any point in multi-dimensional parameter space using a fast interpolation scheme.
This greatly reduces the discrepancy between the model photometry and the data arising from the coarseness of a model grid, obviating
the need for complicated template error correction terms with functional forms that are hard to motivate.
The interpolation also enables continuous probability density functions to be used as priors on the parameters,
which may be applied individually for each parameter or jointly on a number of parameters together in the current implementation of the code. 
 
 In fitting the LLAMA AGN, we applied a lognormal prior distribution on \lagn\ using the information
 available from their X-ray luminosities. Adopting the best-fit relationship from \citet{gandhi09}, we calculated
 a 12 \mics\ luminosity from the AGN component from \lx, and extrapolated this to an estimate of \lagn\
 using the mean AGN IR template from \citet{mullaney11} ($R_{160}  = 1.5\times10^{-2}$). We set the mode
 of the prior distribution on \lagn\ to this value and took a fixed standard deviation of 1 dex, which conservatively
 combines the uncertainty on the X-ray--MIR relationship, the errors on \lx, and the range of ratios of the 12 \mics\ luminosity
 to \lagn\ in the family of AGN empirical templates. 
 
 We also used a similar approach to derive an upper limit on \lagn\ in the LLAMA controls sample. A custom analysis following 
 \citet{koss13} with the more sensitive 105 month {\it SWIFT}-BAT survey does not detect any of the inactive galaxies
 to a $2\sigma$ limit of $4.2\times10^{-12}$ \ergs cm$^{-2}$. Adopting a ratio of \lx\ to
 the $14-195$ keV luminosity previously noted for BAT AGN \citep[$\approx 0.4$;][]{ricci17}, we use this limit to calculate a maximum \lx\ that could arise
 from any possible weak X-ray AGN among the control galaxies. Converting this to an equivalent limiting \lagn\ using our shallowest AGN template
 \citep{symeonidis16}, we set a uniform prior distribution on the AGN luminosity of the control sample
 with a very broad span of 10 dex up to this limiting value. We also adopted a uniform prior for $R_{160}$ covering the full range
 shown by the AGN template library, between $2.7\times10^{-3}$ and $4.6\times10^{-2}$.
 
 We applied a broad uniformly-distributed prior on \lgal\ with a span of 10 dex. 
 For each galaxy with at least one detection in a FIR band ($\lambda > 20$ \mics), 
 we computed the geometric mean of the measured monochromatic luminosities in these bands and took this as the central 
 value of the prior distribution. In the galaxies without such detections, we chose a central value of \lgal$=10^{42}$ \ergs. 
We stress that this prior distribution is very uninformative, allowing the likelihood of the model to determine the posterior
distributions of \lgal\ while preventing the code from exploring unphysically low or high values. We adopted a normal prior
for $\alpha_{\rm D}$ with a mean of 2.0 and a dispersion of 1.4, which captures the distribution found among massive star-forming
galaxies in the local Universe \citep[e.g.][]{rosario16}.

While the details of the stellar component of the SEDs are not critical to our results, we nevertheless applied reasonable physical priors for the parameters
of the SSPs. The stellar mass was allowed to vary uniformly over 10 dex centred on $10^{10}$ \msun, while the SSP ages and abundances were
constrained with normally distributed priors centred on 1.0 Gyr and solar abundance respectively, with dispersions of 1.0 dex and a factor of 2 
respectively.

The fitting was performed using the Markov-Chain Monte-Carlo (MCMC) engine EMCEE \citep{foreman13} with its affine-invariant ensemble sampler.
We used a likelihood function that followed the specifications of \citet{sawicki12}, the product of a
standard $\chi^{2}$ likelihood for high S/N flux measurements with (assumed) gaussian errors, and likelihood based on the error function to
incorporate the upper limits in the photometry. We ran 40 MCMC chains with 600 steps and a fixed burn-in phase of 300 steps. The sampling
was dense enough to converge on the joint multi-parameter posterior distributions in fits of all the galaxies, while still permitting each
fit to complete with a sensible processing time of $<30$ minutes.

\subsection{Comparative methodology} \label{distributions}

A thorough understanding of the properties of the cold molecular gas in the LLAMA galaxies requires us
to combine direct measurements of CO fluxes, which generally have well-behaved gaussian errors, with estimates of quantities
such as \lgal, derived from our multi-component SED fits. The uncertainties on the latter can have a distribution that differs considerably from 
normal, due to the complex likelihood space that arises from the SED models. In addition, our controlled experiment investigates the
differences of various quantities between the AGN and their control galaxies, while being aware
of their often substantial uncertainties. The different sensitivities of the FIR photometry between AGN
and inactive galaxies (Section \ref{fir_phot}) adds another layer of intricacy when making these comparative assessments.

In this subsection, we describe the statistical approach we have taken in estimating the distributions of various measured or
modelled quantities for an ensemble of galaxies (AGN or inactive), and in quantifying the differences between 
ensembles of controlled pairs of galaxies. 

We begin by drawing a distinction between the estimate of a quantity $q$ for a particular galaxy $i$, denoted by $p_{i}(q)$, and
the estimate of the distribution of $q$ for an ensemble of galaxies $\bar{D}(q)$. Both are estimations, in the sense that they
represent our knowledge of the truth, in this case the actual value of $q$ for a galaxy ($Q$) and the actual distribution of $q$
for the ensemble ($D(q)$), in the presence of sampling error and the uncertainties of the measured data. For $q \equiv$ \lco,
 $p_{i}(q)$ for a given galaxy is determined by the normally distributed measurement errors on its \coline\ flux 
 and the uncertainty on its luminosity distance (typically 15 per cent for Tully-Fisher based distances). For $q \equiv$ \lgal,
$p_{i}(q)$ is instead determined by the posterior distribution from the Bayesian fitting exercise. When evaluating
functions of estimated quantities (e.g., the gas fraction in Section \ref{gas_fracs} which depends on \lco\ and \smass), 
we use the bootstrap technique\footnote{We use a fixed number of 2000 bootstrap
samples for each evaluation.} to sample $p_{i}(q)$ for each of the independent variables
and evaluate the functional relationship to obtain $p_{i}(q)$ for the derived quantity. In this way, we propagate uncertainties
and covariances consistently for all measured and derived quantities in this study.

A special note about our treatment of limits. For a quantity $q$ that is assessed to have a limiting value $L_{i}$ for a galaxy $i$, 
we assume that $p_{i}(q)$ is a uniform distribution between the $L_{i}$ and $\pm3$ dex of $L_{i}$ (depending
on whether it is an upper or lower limit). This is a very uninformative assumption, designed to treat
the information from the limits as conservatively as possible.

The accuracy of $\bar{D}(q)$ for any particular ensemble of galaxies is set by $p_{i}(q)$ for the individual objects
and sampling error due to the finite size of the ensemble. We use the Kernel Density Estimate (KDE) technique to 
describe $\bar{D}(q)$ for visual assessment in figures, with the choice of uniform bandwidth listed in the associated captions.

While the bootstrap approach outlined above
allows us to describe $p_{i}(q)$ accurately, sample size variance is built into the design of the LLAMA experiment and cannot
be easily overcome. While LLAMA is one of the largest host galaxy-controlled AGN surveys with uniform millimetre spectroscopy,
the subsample size of 18 objects still places severe limitations on the discrimination of fine differences between AGN and
their control galaxies. 

We perform Kolmogorov--Smirnov (K--S) tests on pairs of bootstrapped samples
of the AGN and inactive galaxies to describe the differences between the $\bar{D}(q)$ of each subsample. These
differences are represented by the median value of P$_{\rm KS}$,
the probability that the two distributions are drawn from a common parent distribution (the Null hypothesis of the test).
Following our choice of significance threshold,  a median P$_{\rm KS} < 0.05$ indicates that the distributions
from the subsamples are statistically distinct.

We also consider the distributions of the difference of a certain quantity (typically expressed as a logarithmic difference) 
between an AGN and its control galaxy, using the ensemble of controlled pairs. These distributions are better
at revealing finer differences between the subsamples, since they factor out systematic variations correlated
with the galaxy properties used for the matching of controls in LLAMA.  For example, \lgal\ is low in early-type galaxies
and high in late-types. The difference distributions only compare AGN and inactive galaxies with the same
morphology (within the matching tolerance); therefore, it serves as a better indicator of potential
morphology-independent differences of \lgal\ between AGN and their controls than a comparison of the separate
\lgal\ distributions of AGN and inactive galaxies, since the latter approach is more affected by morphology-dependent
scatter within each subsample. 

We construct difference distributions using the bootstrap approach. In each realisation, we randomly select one control
galaxy for each AGN, yielding 18 independent control pairs per realisation. This makes full use of the additional
information available for AGN with multiple control galaxies, while maintaining equal statistical weight for all AGN.
We note, however, that our results do not strongly differ if we took an approach that uses all control pairs in each
bootstrap realisation to construct difference distributions. We describe these distributions using KDE for visual purposes, 
and derive a median difference, its uncertainty, and the variance of the differences from the 
bootstrapped samples. 

We also account for sampling error by performing a one-sample Student's T-test on each bootstrap realisation, 
which determines whether it differs significantly from a zero-mean normal distribution. A median value of
P$_{\rm T} < 0.05$ implies that the difference distribution has a mean value that is significantly offset from zero. 
We do find any statistically significant offsets between AGN and control for any of the quantities studied in this work. 
As a guide for future studies, we state the minimum difference that we can significantly measure with the
size of the LLAMA sample using simulations based on the observed difference distributions.

Key information is displayed in the distinctive three panel plots used throughout Section \ref{results} and
in Table \ref{stats_table}.

\input{Table3.tex}

\section{Results} \label{results}

\subsection{\coline\ intensities} \label{co_lums}

\begin{figure}
\includegraphics[width=\columnwidth]{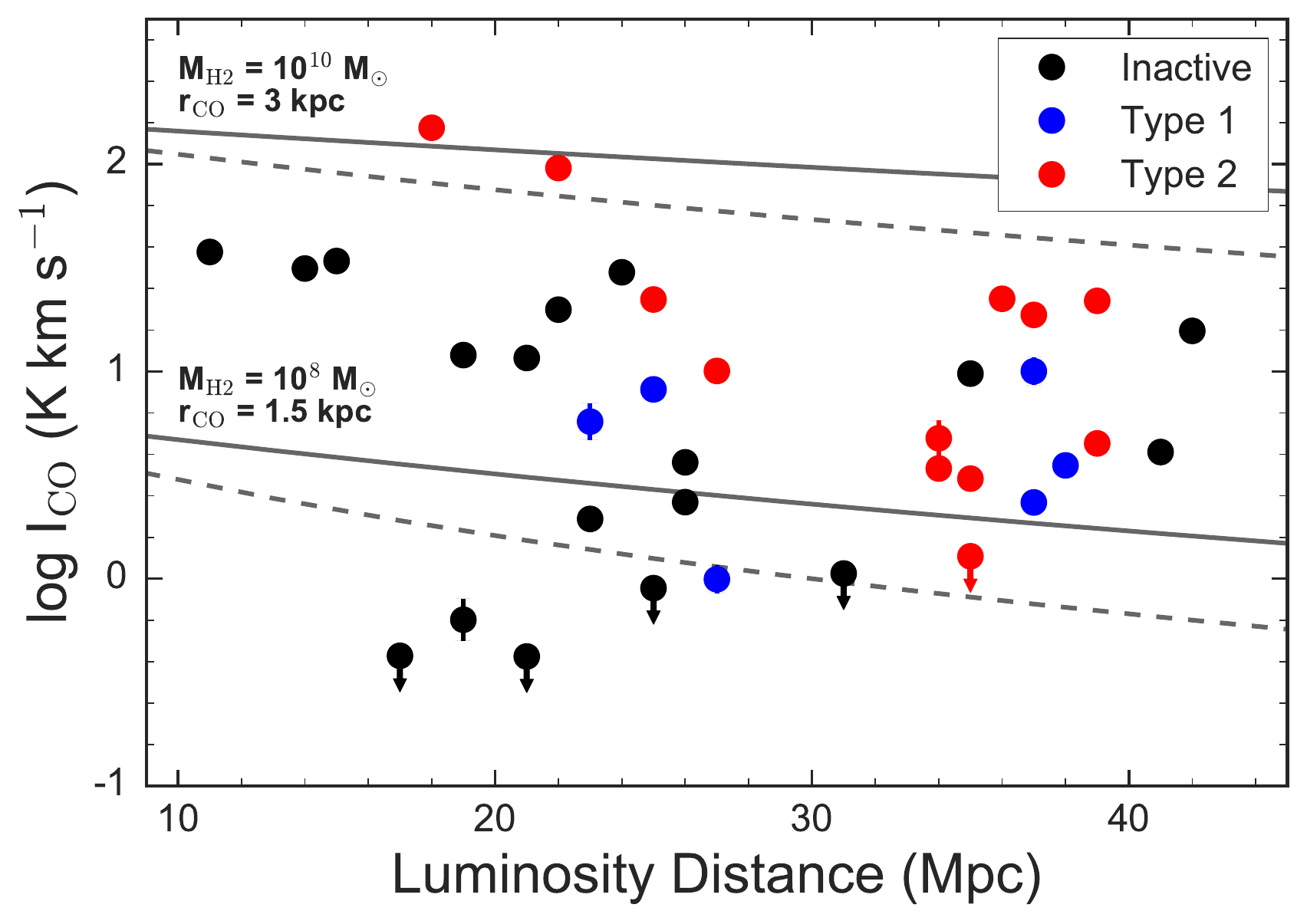}
\caption{The observed velocity-integrated \coline\ surface brightness (\ico) of the 
LLAMA galaxies plotted against their luminosity distances. AGN are shown
as blue (Type 1) and red (Type 2) points, and inactive galaxies are shown as black points. 
Arrowheads depict $3\sigma$ CO upper limits. The tracks describe the change in \ico\
of smooth exponential model discs with distance, if observed with a single dish telescope
with the APEX beam. We show the expectations for two models with molecular Hydrogen
masses (M$_{H2}$) of $10^{8}$ \msun\ (lower tracks) and $10^{10}$  \msun\ (upper tracks) and disc
scale lengths (r$_{\rm CO}$) of 3 kpc and 1.5 kpc respectively. 
Solid and dashed lines represent discs that are face-on and inclined
at $70^{\circ}$ with respect to the line of sight.  
}
\label{lco_distance}
\end{figure}

In Figure \ref{lco_distance}, we plot the velocity-integrated \coline\ surface brightness (\ico) 
against distance for the LLAMA galaxies, splitting them into the AGN (coloured points)
and inactives (black points). The AGN are further distinguished based on their optical classification: 
Seyfert 1--1.5 are marked as Type 1 (AGN with low levels of optical extinction), and 
Seyferts 1.8 and higher (including Seyferts 1i and 1h) as Type 2 (AGN with moderate to high levels of optical extinction).
The LLAMA control galaxies are typically closer than the LLAMA AGN, with a median distance difference
of 12 Mpc.  

As a guide to the eye, the lines in Figure \ref{lco_distance} show the expected variation of \ico\ with distance for smooth exponential discs 
of a given molecular gas mass, observed with the APEX beam modelled as a Gaussian with width $\sigma_{\rm b}= 13\farcs55$). 
We plot tracks for two fiducial disc models with different sizes and total molecular Hydrogen mass (M$_{H2}$), at two different inclinations
with respect to the line of sight. Due to the radial gradient in the CO emission in discs, the observed beam-averaged surface brightness
drops slowly with distance. A fair comparison of the CO intensities of LLAMA AGN and control should account for the systematic change
in \ico\ with distance expected in galaxy discs. We derive a correction to \ico\ based on smooth exponential disc models
to allow this comparison.

Following \citet{leroy08}, we estimate the exponential scale length of the CO disc in each galaxy from the size of the stellar disc:

\begin{equation}
r_{\rm CO} = 0.2 \times r_{25}
\end{equation}

\noindent where $r_{25}$ is the galaxy's $B$-band isophotal radius at 25 mag arcsec$^{-2}$ (Table \ref{basic_data}). 
This relation has a scatter of $\approx 35$\%.
As discussed in \citet{lisenfeld11} and \citet{boselli14}, r$_{\rm CO}$ and the inclination of the exponential molecular gas disc (assumed to be the
same as the inclination of the stellar disc) together define a CO profile which may be windowed by the single-dish telescope 
beam and integrated to estimate the observed CO flux of an observation of the galaxy. We calculate the ratio of this
integral for the actual distance of the galaxy and assuming that it is at a reference distance of 25 Mpc. Multiplying the observed \ico\
by this ratio, we calculate the expected CO surface brightness at the reference distance (\icor), correcting for first-order distance-dependent systematics.

\begin{figure}
\includegraphics[width=\columnwidth]{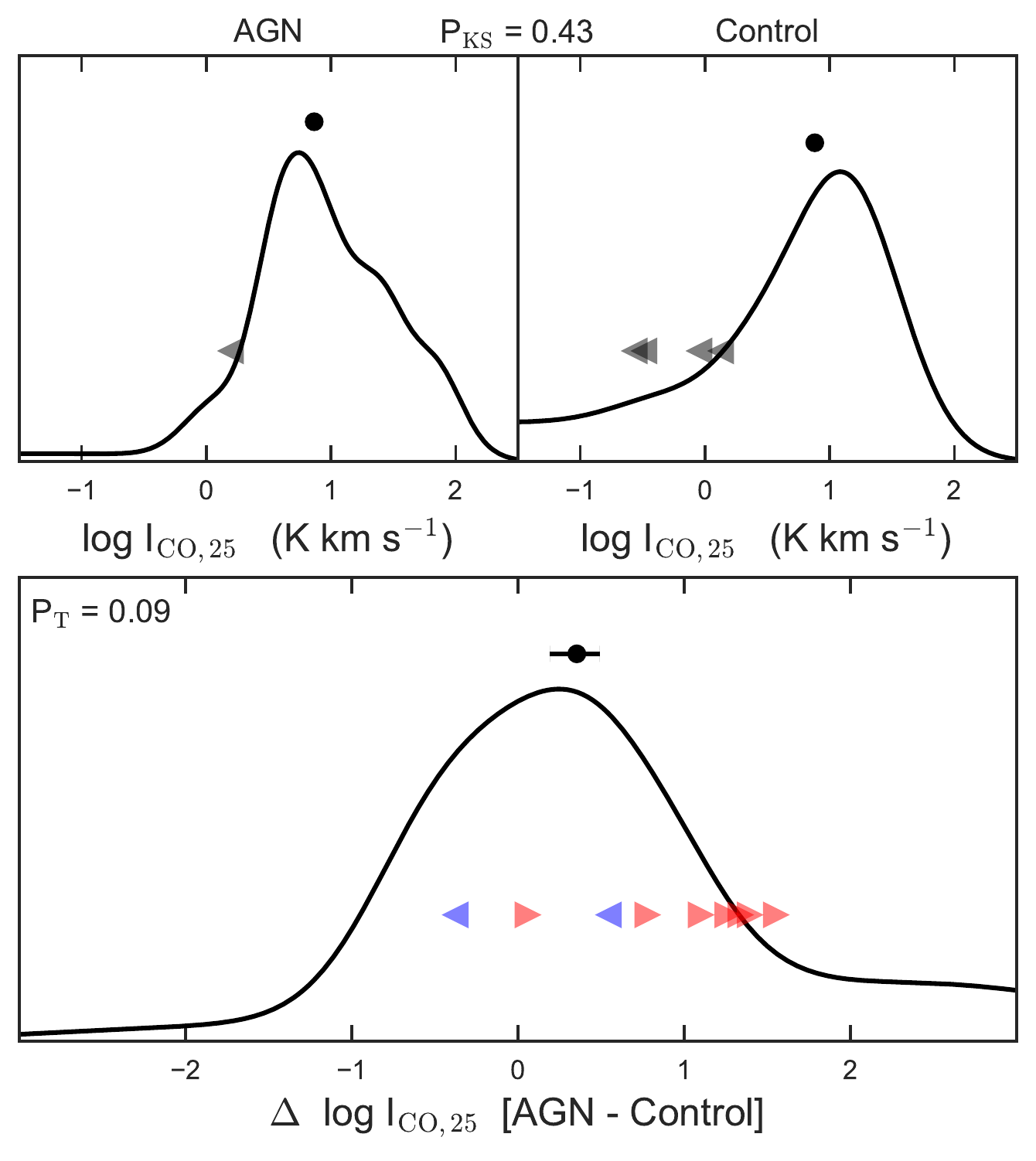}
\caption{ {\bf Upper panels:} Distributions of the \coline\ line intensities (\icor) of the LLAMA AGN (left) and inactive galaxies (right),
referred to a fixed distance of 25 Mpc and scaled to the APEX beam.
A gaussian kernel-density estimator (KDE) with a bandwidth of 0.3 is used to display these distributions. 
Nominal $3\sigma$ upper limits are shown as left-facing arrows, but this information is also folded into the distributions. 
From 2000 random samples obtained by bootstrapping, 
we obtain the medians of the distributions and their uncertainties (full black symbols with error bars) and the
median two-sample K--S probability (P$_{\rm KS}$). 
{\bf Lower panel:} The logarithmic difference ($\Delta$) of \icor\ for matched pairs of AGN and control galaxies. The median $\Delta$
is shown as a full black symbol, with uncertainties from 2000 random bootstrapped samples. 
Nominal upper/lower limits on $\Delta$ are shown as left/right-facing arrows, but this information is also folded into the distribution.
The median probability of a Student's T-test of zero-mean (P$_{\rm T}$) is obtained from the bootstrapped samples.
}
\label{lco_dists}
\end{figure}

In the top two panels of Figure \ref{lco_dists}, we compare the distributions of \icor\ of the AGN and the inactive
galaxies. Their similar medians and a high value of P$_{\rm KS}\approx 45$ per cent indicates that the distributions are 
statistically indistinguishable. 

In the lower panel, we consider the distribution of the logarithmic difference of \icor\ for the controlled pairs of AGN and inactive galaxies. 
While we show the limits for visual inspection purposes, their effects are taken into account when estimating the distribution 
in the fashion outlined in Section \ref{distributions}. The overall distribution of differences is wide, and has a median value of $0.36\pm0.15$ dex, 
implying \coline\ intensity that is $\times 2.5$ higher in the AGN than in their matched inactive galaxies (at a significance of $\approx 2\sigma$).
However, we cannot rule out that a difference at this level is solely due to stochasticity between the two subsamples, as the Student's T probability
is $>5$\%. Simulations allow us state that if the mean \ico\ of the AGN and control galaxies differed by a factor of $> 4$, we would have seen this
effect clearly in our sample.

\subsection{Gas fractions} \label{gas_fracs}

\begin{figure}
\includegraphics[width=\columnwidth]{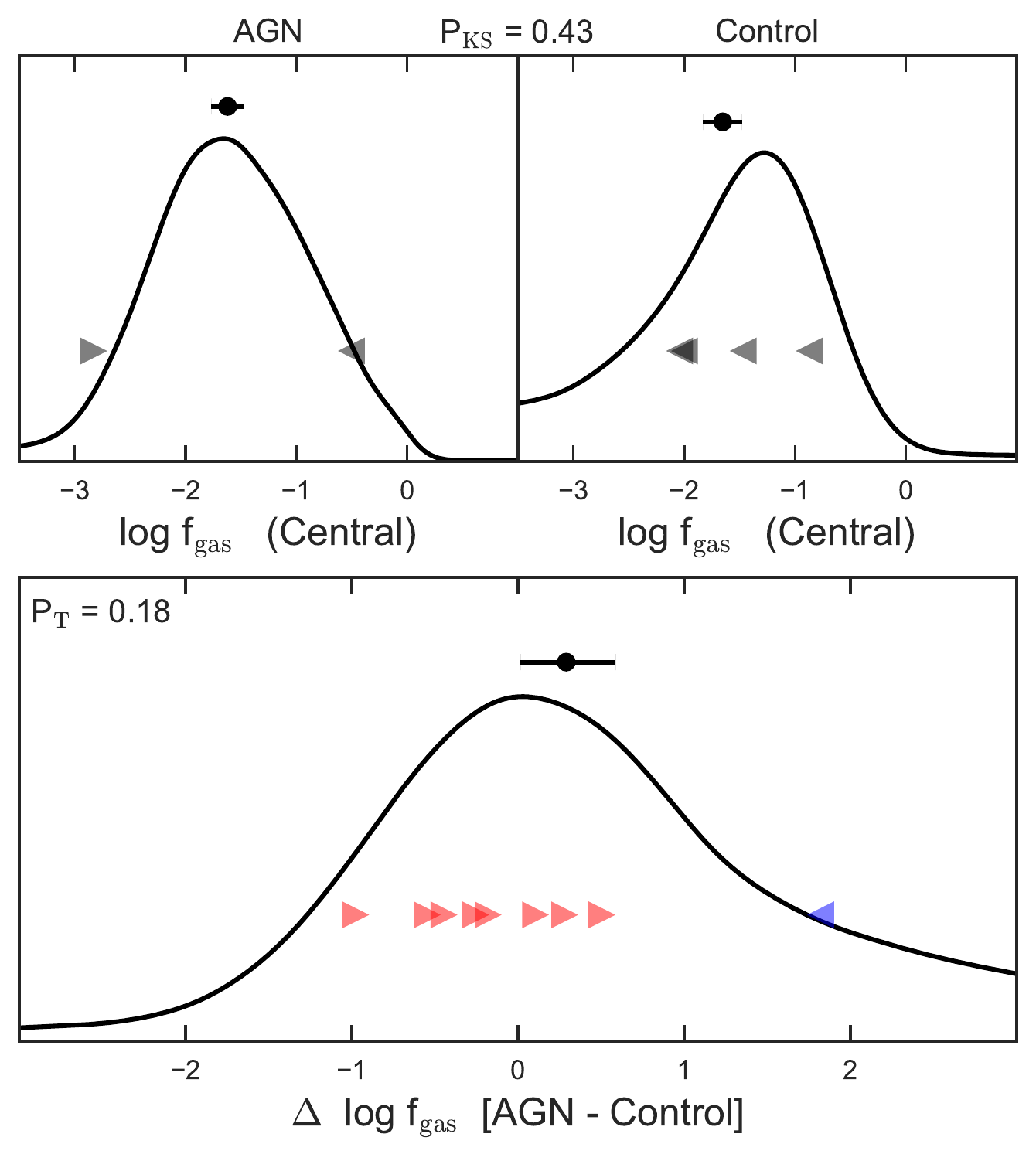}
\caption{ {\bf Upper panels:} Distributions of the central gas fraction ($f_{\rm gas}$) of the LLAMA AGN (left) and inactive galaxies (right).
A gaussian kernel-density estimator (KDE) with a bandwidth of 0.2 is used to display these distributions. 
{\bf Lower panel:} The logarithmic difference ($\Delta$) of $f_{\rm gas}$ for matched pairs of AGN and control galaxies. See the caption of
Figure \ref{lco_dists} and Section \ref{distributions} for details of the elements of this figure.
}
\label{gasfrac_dists}
\end{figure}

From the beam-matched photometry of their 2MASS images, we estimate the stellar masses ($M_{\star}$) of the LLAMA galaxies within the regions
covered by the CO observations and use these to derive their central molecular gas fractions. 

We obtain the central $M_{\star}$ of our targets by multiplying their 2MASS K$_{s}$ beam-matched luminosity 
by an appropriate K-band mass-to-light (M/L$_{K}$) ratio.
We examined a suite of composite population synthesis models from the \citet{maraston05} library and established that
stellar populations with ages $\gtrsim$ Gyr, such as those found in the centres of massive galaxies, show a range of M/L$_{K}$ with a typical
value of 3.2 and scatter of 0.3 dex. We therefore considered a lognormal distribution of M/L$_{K}$ specified by these
values in our bootstrap analysis. The scatter of M/L$_{K}$ is the dominant source of uncertainty in these mass 
estimates, being greater than both the errors of the 
photometric measurements and the errors on nuclear luminosities from \citet{burtscher15}.

The CO luminosities (\lco) of our targets are calculated following \citet{solomon05}:

\begin{equation}
L^{\prime}_{\rm CO} = 3.25\times10^{7}  \, \frac{S_{\rm CO} \, R_{12} \,D_{L}^{2} }{(1+z)^{2} \times 230.54\, {\rm GHz}}  \; \; \textrm{K km s}^{-2} \textrm{pc}^{2}
\end{equation} 

\noindent where $S_{\rm CO}$ is the velocity-integrated flux of the \coline\ line in Jy km s$^{-1}$, D$_{L}$ is the luminosity distance in 
Mpc and $z$ is the redshift. R$_{12}$ is the  CO 1$\rightarrow$0/CO 2$\rightarrow$1  brightness temperature ratio which refers \lco\ to the 
CO 1$\rightarrow$0 line as is the custom in most studies. In this work, we assume a fiducial value of  R$_{12} = 1.4$ \citep{sandstrom13}.

We adopt a CO-to-H$_{2}$ conversion factor \alphaco\ $= 1.1$ \msun pc$^{-2}$/(K km s$^{-2}$) (referred to CO 1$\rightarrow$0). 
This is lower than
the canonical value for the Milky Way disk \citep[\alphaco\ $ = 4.4$ \msun pc$^{-2}$/(K km s$^{-2}$);][]{dame01, bolatto13}, 
but is characteristic of the central kpc of metal-rich galaxies, where \alphaco\ is known to be somewhat depressed \citep{sandstrom13}. 
The average \alphaco\ can vary by $\times2$, and be lower by as much as a factor of five in some systems.
This drop is usually attributed to a change in the properties of the cold molecular clouds in galactic centers, due to increased
turbulence in the clouds or greater pressure in the circum-cloud ISM. There remains, however, no clear association 
between variations in \alphaco\ and the presence of AGN in galaxies. We adopt an \alphaco\
uncertainty of 0.3 dex for our main analysis, and we comment on the effects of potentially lower values when discussing
our findings in Section \ref{discussion}.

The molecular gas fraction is defined as:

\begin{equation}
f_{\rm gas}  \; =  \; \frac{M_{H2}}{M_{H2} + M_{\star}} \\
\end{equation}

\noindent where M$_{H2}$ = \alphaco$\times$\lco\ is the mass in H$_{2}$ in solar units as with $M_{\star}$.

Figure \ref{gasfrac_dists} shows the distributions of $f_{\rm gas}$ of the AGN and inactive galaxies, folding in
the uncertainties and systematics described above.  As we found for the CO line surface brightnesses, the relative abundance of molecular gas
in AGN hosts is indistinguishable from similar inactive galaxies: the medians of the distributions are within 0.1 dex and 
P$_{\rm KS}$ $\approx 45$ per cent. The distribution of the logarithmic difference in the gas fraction between AGN and control
galaxies peaks at zero, with a slightly positive median value ($0.3\pm 0.3$) because the numerous lower limits on $f_{\rm gas}$
lead to a longer positive tail to the estimated distribution. 
We conclude that our AGN hosts do not show any conclusive evidence for either an enhanced or depressed
incidence of cold molecular gas relative to stars in their central regions. With the size of the LLAMA sample, we would
have been sensitive to differences greater that a factor of 5.5.

\subsection{Integrated IR luminosities} \label{ir_luminosities}

\begin{figure}
\includegraphics[width=\columnwidth]{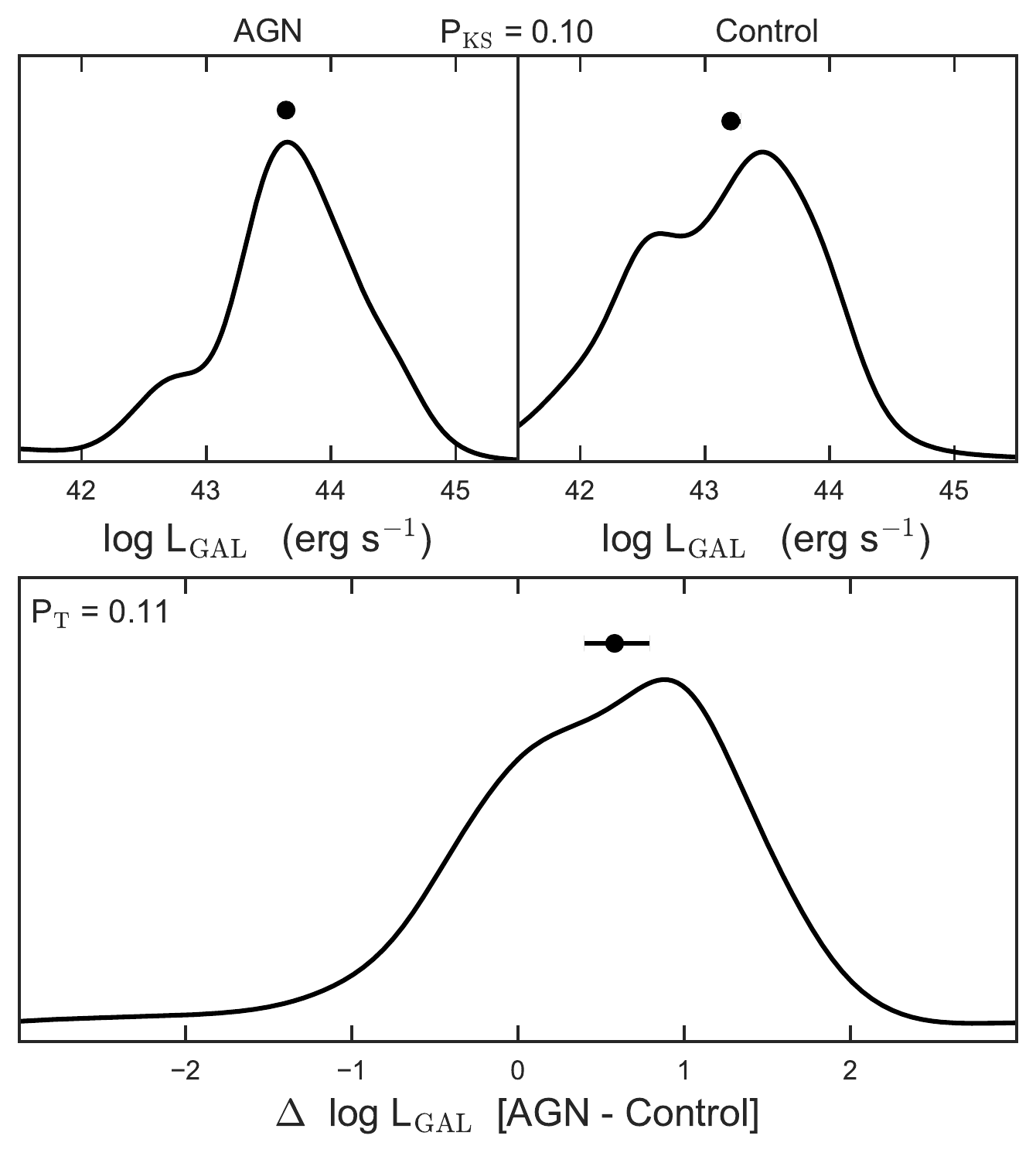}
\caption{ {\bf Upper panels:} Distributions of the integrated IR luminosity from the galaxy (\lgal) for LLAMA AGN (left) and inactive galaxies (right)
derived from SED fits. A gaussian kernel-density estimator (KDE) with a bandwidth of 0.3 is used to display these distributions.
{\bf Lower panel:} The logarithmic difference ($\Delta$) of \lgal\ for matched pairs of AGN and control galaxies. See the caption of
Figure \ref{lco_dists} and Section \ref{distributions} for details of the elements of this figure.
}
\label{lir_integ_dists}
\end{figure}

The SED fitting procedure allows us to decompose the galaxy-integrated IR luminosities of the LLAMA AGN into
emission from AGN (\lagn) and emission from extended cold dust that is primarily heated by star-formation (\lgal). 
The inactive galaxies are also fit in the similar fashion, with a cap on \lagn\ on the grounds that none show signs
of nuclear activity. 

In the top panels of Figure \ref{lir_integ_dists}, we compare the marginalised distributions of 
\lgal\ of the AGN and inactive galaxies taken directly from the outputs of our SED fits.
The median \lgal\ of the AGN is 0.4 dex higher than the inactive galaxies, but this could arise due to sampling error,
as a K--S test indicate that the \lgal\ distributions are not significantly indistinct.
The distribution of logarithmic difference in \lgal\ from the control pairs shows a
positive shift ($0.6\pm0.2$ dex) towards a higher IR luminosity from AGN hosts
after controlling for galaxy mass, inclination and Hubble Type. This offset is below our adopted
significance threshold from the T-test, but we can exclude differences between AGN and control populations that
are greater than a factor of 5.5.

\subsection{Beam-matched IR luminosities and central SFR}

\begin{figure}
\includegraphics[width=\columnwidth]{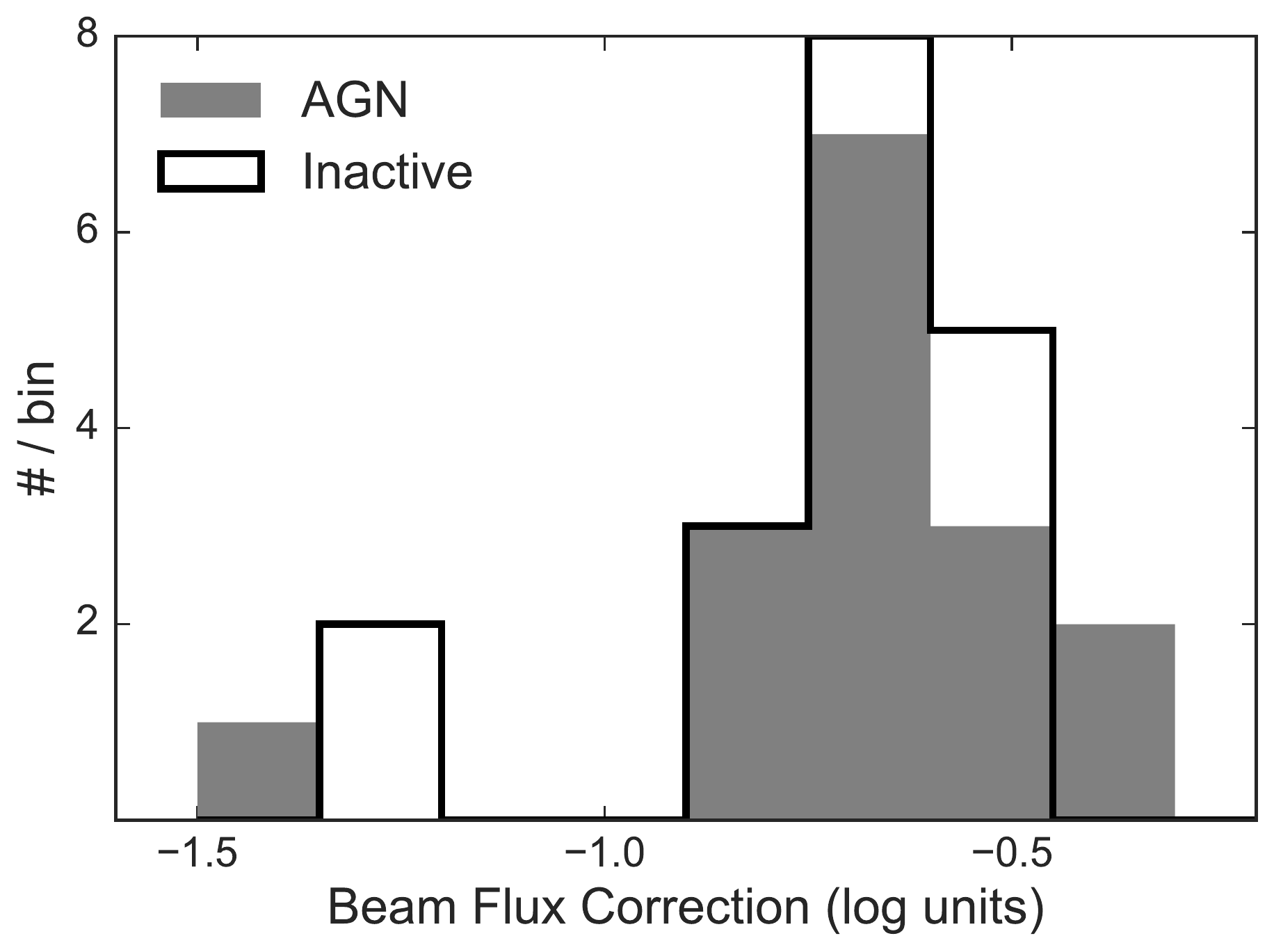}
\caption{Distributions of the CO telescope beam matching factor ($R_{ir}$) applied to the integrated IR luminosity from the galaxy (\lgal)
to derive central IR luminosities and SFRs. Separate distributions are shown for LLAMA AGN (filled histogram) and inactive galaxies (open histogram).
These distributions are broadly similar.
}
\label{beam_correction}
\end{figure}

In Sections \ref{mir_phot} \& \ref{fir_phot}, we developed the approach taken to measure {\it WISE} 22 \mics\ and {\it Herschel}/PACS 160 \mics\
photometry matched to the beam of the CO spectroscopic observations. The images in these bands offer the highest available
spatial resolution at wavelengths dominated by the thermal infrared emission from star-formation, 
a necessary handle on the central SFRs in these galaxies.
 
Two AGN  (NGC 4593 \& ESO 021-G004) lack {\it Herschel} imaging; for these objects
we do not possess a reliable measure of the resolved thermal dust emission, 
since the {\it WISE} 22 \mics\ band is strongly dominated by AGN light. 
We take their integrated 
\lgal\ as a conservative upper limit to their central FIR luminosities. 

The ratio of the beam-matched flux to the integrated flux of a galaxy in its respective band (hereafter $R_{ir}$)
gives us the scaling to convert the galaxy's integrated \lgal\
to an estimate of the central \lgal. In Figure \ref{beam_correction}, we compare $R_{ir}$ for the AGN and inactive
galaxies. Despite the differences in the bands used to derive this ratio, its distributions
are similar in the two subsamples, which suggests that there are no large differences
in the average dust emission profiles from star-formation between the AGN and inactive galaxies. 

We use a modified form of the relationship between SFR and \lgal\ to derive the central SFR, as follows:
 
\begin{equation}
\label{sfr_cal}
SFR   \; =   \; \frac{R_{ir} \, L_{\rm gal}}{4.48\times10^{43}}  \; \; \textrm{M}_{\odot}  \textrm{ yr}^{-1}
\end{equation}

\noindent which is based on the calibration from \citet{kennicutt12}, in the limit that all the radiation from
the young stars of a continuous star-forming population with a Chabrier IMF is absorbed by dust \citep{rosario16}.
The uncertainty on this calibration is approximately 0.5 dex, which will dominate the error budget
of this calculation over any uncertainty in $R_{ir}$ or differences in the dust SEDs of the integrated and central
dust emission.

\subsection{Molecular gas scaling laws and SF depletion times} \label{sf_efficiency}

Molecular gas is the raw material for star formation. The SFR associated with a unit mass 
of molecular gas (the star-formation efficiency or `SFE') provides fundamental insight into the physical conditions of
the molecular phase and the environments of star-forming regions within it. The mass of molecular gas is related to the luminosity
of the CO line (\lco), and the SFR determines the majority of the IR luminosity (\lgal) of star-forming galaxies. 

\begin{figure*}
\includegraphics[width=0.8\textwidth]{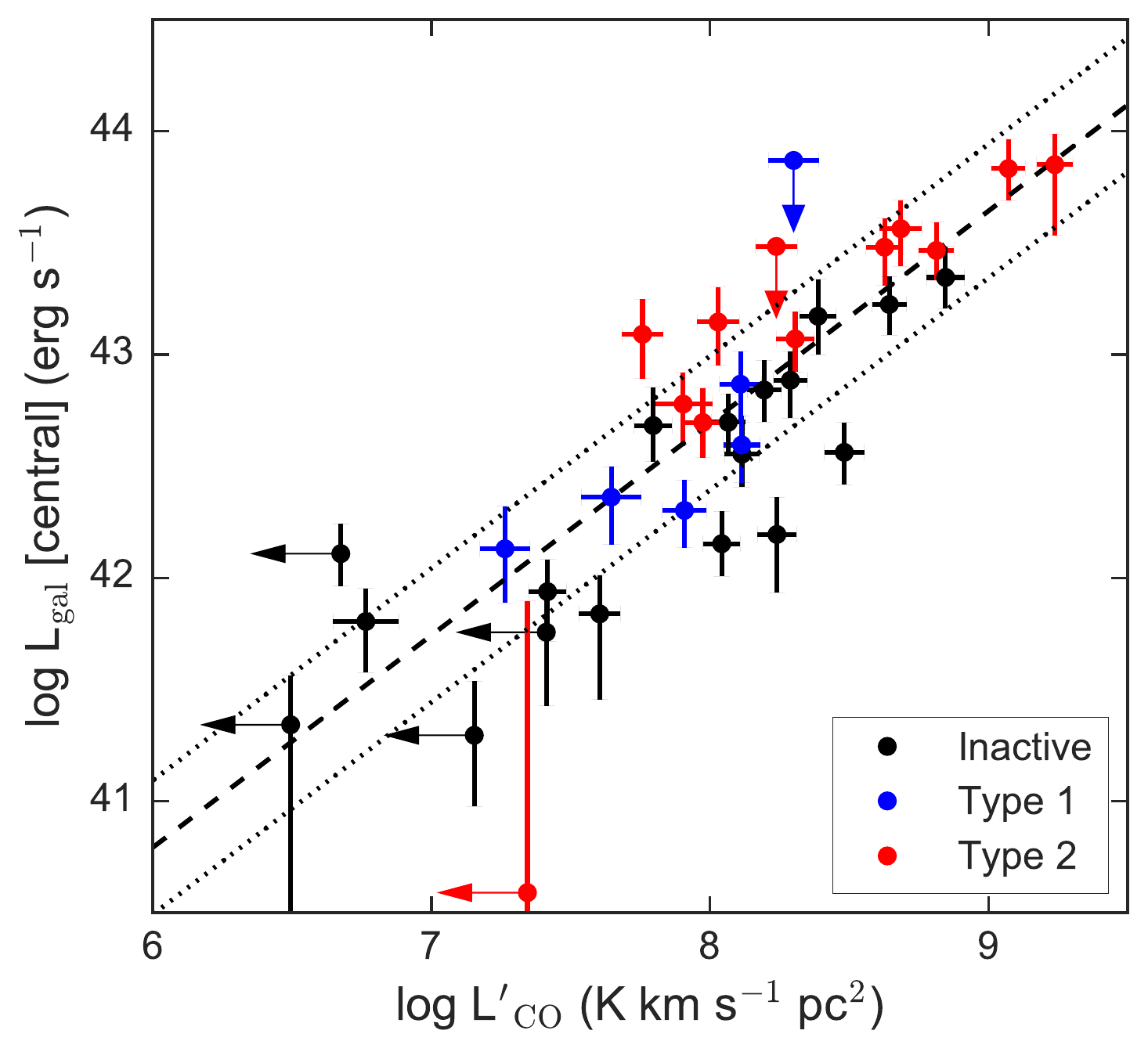}
\caption{The \coline\ luminosities (\lco) of LLAMA sources plotted against their central (beam-matched) IR luminosities.
AGN are shown as coloured points, differentiated into Type 1s (blue points) and Type 2s (red points), while inactive
galaxies are shown with black points. The dashed line shows the relation from \citet{leroy13} for the disks of normal
star-forming galaxies in the HERACLES survey, converted to the plotted quantities using calibrations used in the present work
(except for \alphaco; see text for details). The dotted lines show the typical 0.3 dex scatter about this relation.
}
\label{lco_lgal_cen}
\end{figure*}

A popular diagnostic of the SFE is the correlation between \lco\ and \lgal: a region of a galaxy
with more efficient star formation will produce a higher \lgal\ for a given \lco\ (if \alphaco\ remains fixed and
\lgal\ is a bolometric measure of the SFR). We plot this correlation in 
Figure \ref{lco_lgal_cen}, differentiating between the two types of
AGN (Type 1/2; Section \ref{co_lums}) and the inactive galaxies. The dashed line in the Figure shows the relationship among 
nearby resolved disk galaxies in the HERACLES survey \citep{leroy13}, 
scaled to a fiducial region of a galaxy that spans 2 kpc\footnote{The relationship given as Eqn.~9 of 
\citet{leroy13} is in terms of surface densities of gas and SFR. For consistency with our approach, we adopt their FUV+24 \mics\ SFR relationship, and
convert to \lco\ and \lgal\ using our own preferred calibrations. We however retain their choice of \alphaco\ which is appropriate for galaxy disks.}. 
The dotted lines delineate the estimated 0.3 dex scatter about this relationship.

Most AGN  (16 of 17) and inactive galaxies (10 of 14) with CO detections lie within the 
expected relationship for normal galaxy disks, implying that the SFEs of the central gas of these systems is typical. 
Therefore, to first order, the AGN and inactive galaxies as a population have similar normal SFEs.
A few inactive galaxies scatter below the relationship, while the AGN tend to occupy the upper envelope of the data points.
A small offset of $\approx 0.3$ dex is measurable between the two subsamples, such that the AGN show a higher 
\lgal for a given \lco, and therefore, a higher overall SFE. This offset is, however, not significant due to the restrictions of the sample
size of LLAMA. It should be pursued with a larger survey with more statistical power.

We examine the central SFE in more detail 
by examining the distributions of the  
molecular gas depletion time \taud, the ratio of M$_{H2}$ to the SFR (in surface densities or integrated over the CO telescope beam). 
This is the lifetime of the molecular gas at a constant level of star-formation; more efficient star-forming environments show shorter depletion times.

\begin{figure}
\includegraphics[width=\columnwidth]{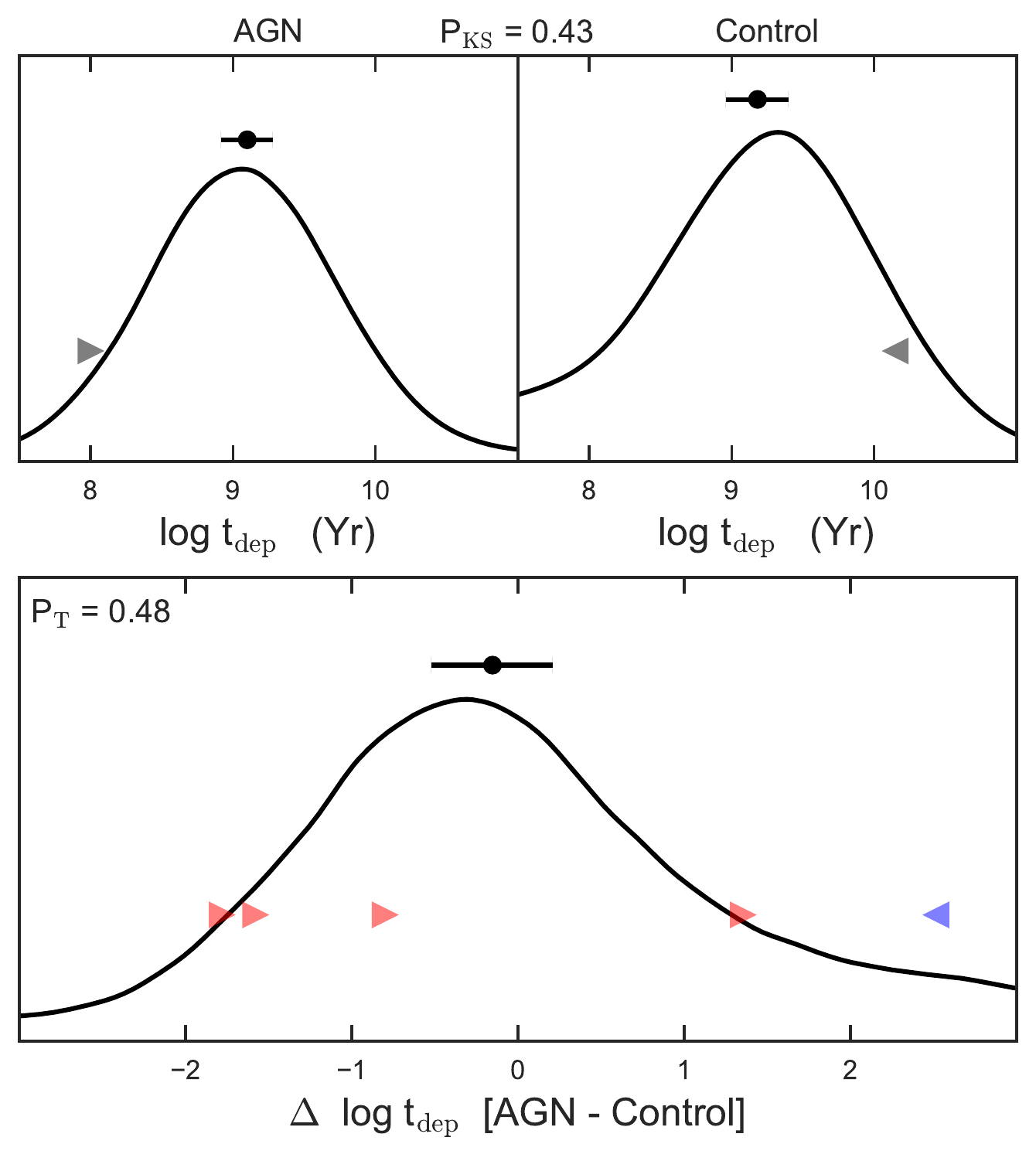}
\caption{ {\bf Upper panels:} Distributions of the central molecular depletion times (\taud) for LLAMA AGN (left) and inactive galaxies (right).
A gaussian kernel-density estimator (KDE) with a bandwidth of 0.2 is used to display these distributions. 
{\bf Lower panel:} The logarithmic difference ($\Delta$) of \taud\ for matched pairs of AGN and control galaxies. See the caption of
Figure \ref{lco_dists} and Section \ref{distributions} for details of the elements of this figure.
}
\label{tau_central_dists}
\end{figure}

We compare the distributions of \taud\ between AGN and inactive galaxies in the upper panels of Figure \ref{tau_central_dists}.
The two subsamples taken individually show statistically similar central depletion times (P$_{\rm KS} \approx$ 43 per cent), 
with median values of slightly over 1 Gyr. This is typical of galaxy disks \citep{leroy08}, confirming the impressions from Figure \ref{lco_lgal_cen}.
The logarithmic difference of \taud\ between the AGN and their control shows a broad distribution that is statistically consistent with a zero offset.
Simulations indicate that our experiment would have confirmed offsets greater than a factor of 6.
We conclude that the central molecular gas properties of the AGN and control galaxies in LLAMA are statistically similar.

\subsection{Molecular gas and AGN properties} \label{agn_corr}

\input{Table4.tex}

Here we explore whether the power output or nuclear environment of AGN is related to the gas content in their central kpc. Using X-ray spectral
fits \citep{ricci17}, we test for correlations between the intrinsic X-ray luminosity (\lx) or the Hydrogen column density of nuclear absorption (N$_{\rm H}$)
against the \coline\ intensity (\icor) corrected for distance-dependent systematics (Section \ref{co_lums}). 
This element of our study does not require the control sample, so we are free to expand our sample of AGN to cover a larger range in AGN power
to improve statistics and the dynamic range available to test for correlations. Therefore, we supplement the LLAMA AGN with 
a set of \coline\ measurements from the literature for eight other {\it SWIFT}-BAT selected AGN at $D_{\rm L} < 40$ Mpc. 
We retain the LLAMA distance limit to curtail Malmquist bias and the need for large distance-dependent corrections. 
The \coline\ measurements of the supplementary AGN, adopted distances, X-ray spectral properties, and galaxy structural properties 
relevant to the \ico\ correction are shown in Table \ref{supplementary_agn}.
Their X-ray luminosities overlap with, but are typically lower, than those of LLAMA. Two heavily X-ray obscured supplementary AGN only have lower limits 
on N$_{\rm H}$.

\begin{figure}
\includegraphics[width=\columnwidth]{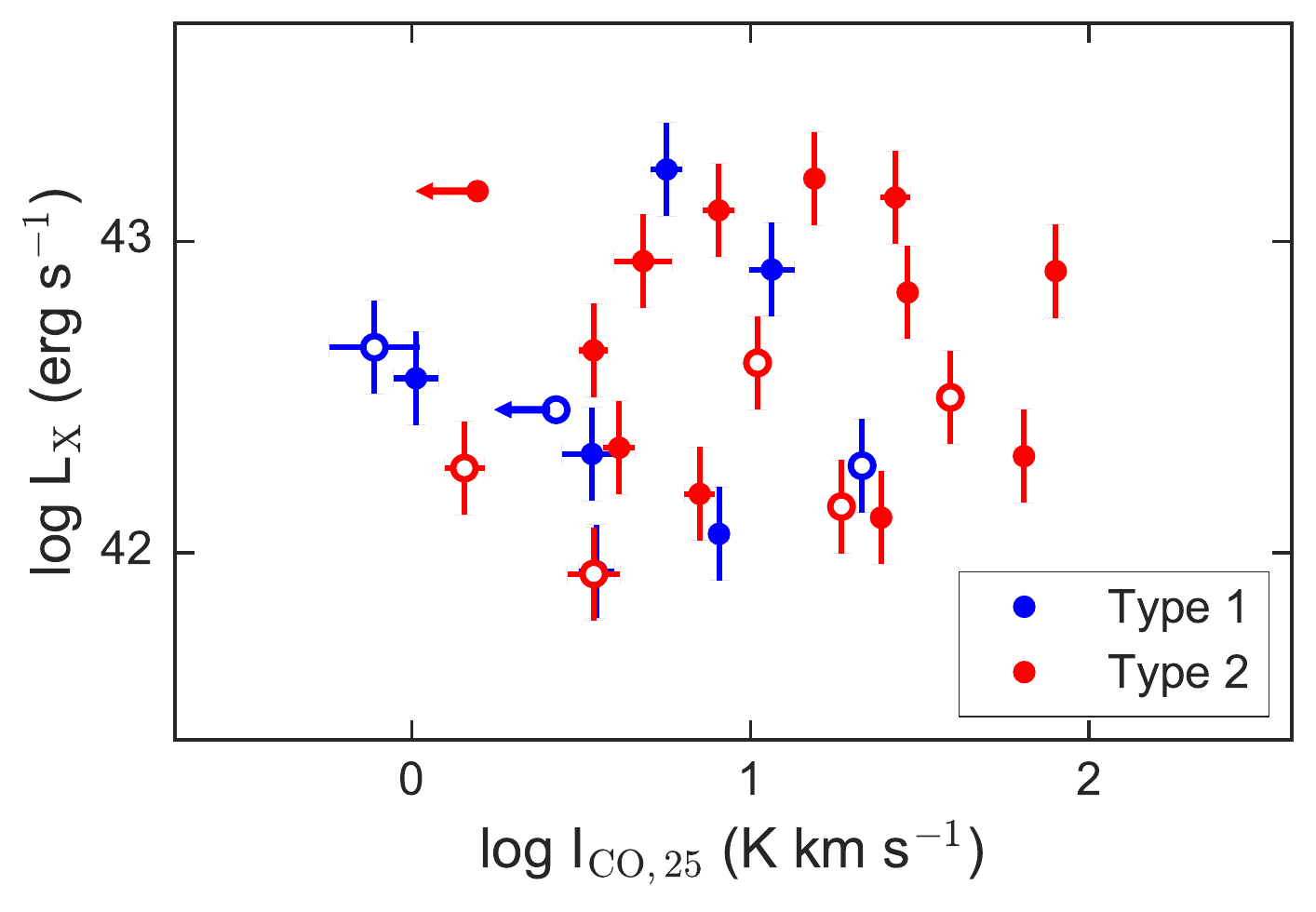}
\caption{The velocity-integrated \coline\ surface brightness corrected to a reference distance of 25 Mpc (\icor; see Section \ref{co_lums} for details) 
plotted against the intrinsic (absorption-corrected) 2-10 keV X-ray luminosity (\lx). Type 1(2) AGN are differentiated by blue (red) points.
LLAMA AGN are plotted with solid points and the supplementary AGN with open points.
}
\label{lco_lx}
\end{figure}

We find no correlation between \icor\ and the power of the AGN measured by \lx\ (Figure \ref{lco_lx}).
A Spearman rank correlation test yields a coefficient of 0.06 and a probability of 78\% that these properties are completely uncorrelated.
This is not surprising given the vastly different scales spanned by the single-dish beams and the accretion disk: 
the amount of cold molecular gas over $\sim$kpc is weakly coupled to the instantaneous 
nuclear accretion rate.

\begin{figure}
\includegraphics[width=\columnwidth]{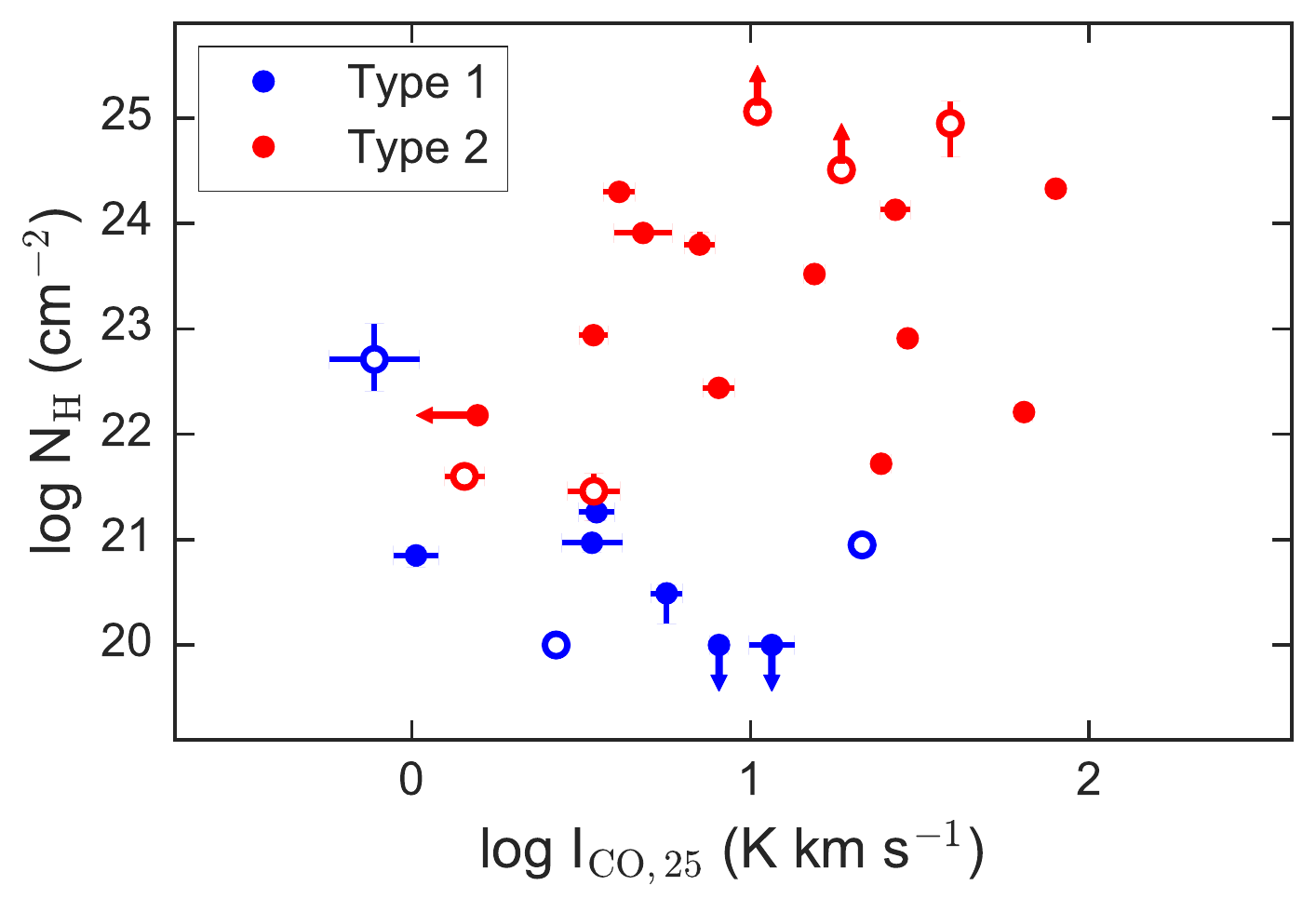}
\caption{The velocity-integrated \coline\ surface brightness corrected to a reference distance of 25 Mpc 
(\icor; see Section \ref{co_lums} for details) plotted against the Hydrogen column density of nuclear absorption (N$_{\rm H}$)
of the LLAMA AGN. Type 1(2) AGN are differentiated by blue (red) points. 
LLAMA AGN are plotted with solid points and the supplementary AGN with open points.
}
\label{lco_nh}
\end{figure}

On the other hand, we find a mild correlation between \icor\ and N$_{\rm H}$ (Figure \ref{lco_nh}) with a
Spearman rank correlation coefficient of 0.4 and a rejection of the null hypothesis of uncorrelated quantities at the 5\% level.
The ability to uncover this trend is certainly due to the large dynamic range in N$_{\rm H}$ spanned by these local AGN.
N$_{\rm H}$ also correlates well with optical extinction towards the broad-line region and with mid-IR extinction towards the AGN 
torus \citep{davies15, burtscher16}, and this can be seen from the vertical differentiation of blue and red points in the Figure. 
While objects with low \ico\ are a mix of Type 1s and Type 2s, there is a preponderance of Type 2s among the AGN with high CO intensities. 
We will discuss this result in more detail in Section \ref{obsc_trends}.

\section{Discussion} \label{discussion}

\subsection{Feedback energy considerations} \label{feedback_calcs}

The large beam size of the single-dish CO observations give us a census of cold gas within $\sim2$ kpc of the nucleus in the
LLAMA galaxies. Here we explore whether this gas could be influenced by the activity in the nucleus, using broad energy
arguments. 

The accretion energy liberated by the AGN can be estimated from their hard X-ray luminosities, using the bolometric correction 
from \citet{winter12} suitable for local BAT-selected AGN. For the period of a typical accreting phase over which the power
from the SMBH remains roughly constant (t$_{\rm AGN}$),
a certain fraction of the emitted energy will couple dynamically with the molecular gas. This radiation coupling ($\epsilon_{r}$) is very uncertain, 
but theoretically motivated models of AGN feedback suggest a value of at least $\sim 5$ per cent \citep{dimatteo05}. Therefore, for each AGN, we can estimate 
an energy E$_{\rm rad}$, a rough measure of the accretion power that could influence molecular gas in the AGN:  

\begin{eqnarray}
\label{erad}
E_{\rm rad}   & \approx  & \epsilon_{r} \, L_{bol} \, t_{\rm AGN} \\
\log L_{bol} & =  &  1.12 \times \log L_{\rm BAT} \; - \; 4.2 
\end{eqnarray}

\noindent where $L_{\rm BAT}$ is the 14--195 keV luminosity of the AGN. We adopt characteristic values of t$_{\rm AGN} = 10^{6}$ yr  
\citep{hickox14, schawinski15} and $\epsilon_{r} = 0.05$.

The exact form of the coupling could be in multiple forms, such as through direct absorption of radiation by molecular or atomic clouds
{\it in situ}, or through a mechanical working fluid such as a thermal wind or relativistic particles in a jet. 
If E$_{\rm rad}$ is distributed across all the molecular gas in the central regions of the
AGN hosts (i.e., within the telescope beam), we could compare it to the gravitational potential energy of this gas, which gives us a sense of whether
feedback from the AGN should destabilise this mass of gas. The gravitational potential in the central regions of our 
galaxies will be dominated by their baryonic material (stars and gas).
The molecular gas fractions are a few per cent (Figure  \ref{gasfrac_dists}) and HI corrections
in the centres of typical disk galaxies are small \citep[e.g.][]{bigiel08}. Therefore, we can use our central 
stellar mass estimates from the beam-matched K-band luminosities, corrected for AGN light, to approximate the
gravitational potential energy of the molecular gas, as follows:

\begin{equation}
\label{pote}
E_{\rm PE}   \; \approx   \; -\frac{G M_{\star} M_{H2}}{\eta \, R_{\rm beam}} \; (1+f_{\rm gas})
\end{equation}

\noindent where R$_{\rm beam}$ is the radius of the galaxy region covered by the CO beam and $\eta$ is a geometrical
factor that encapsulates our ignorance about the spatial distribution of the molecular gas with respect to the 
gravitational potential of the galaxy. $\eta$ has a value of 1 for gas that is uniformly distributed in a disk of size R$_{\rm beam}$
within a spherical isothermal potential. We take this as our baseline assumption. 

\begin{figure}
\includegraphics[width=\columnwidth]{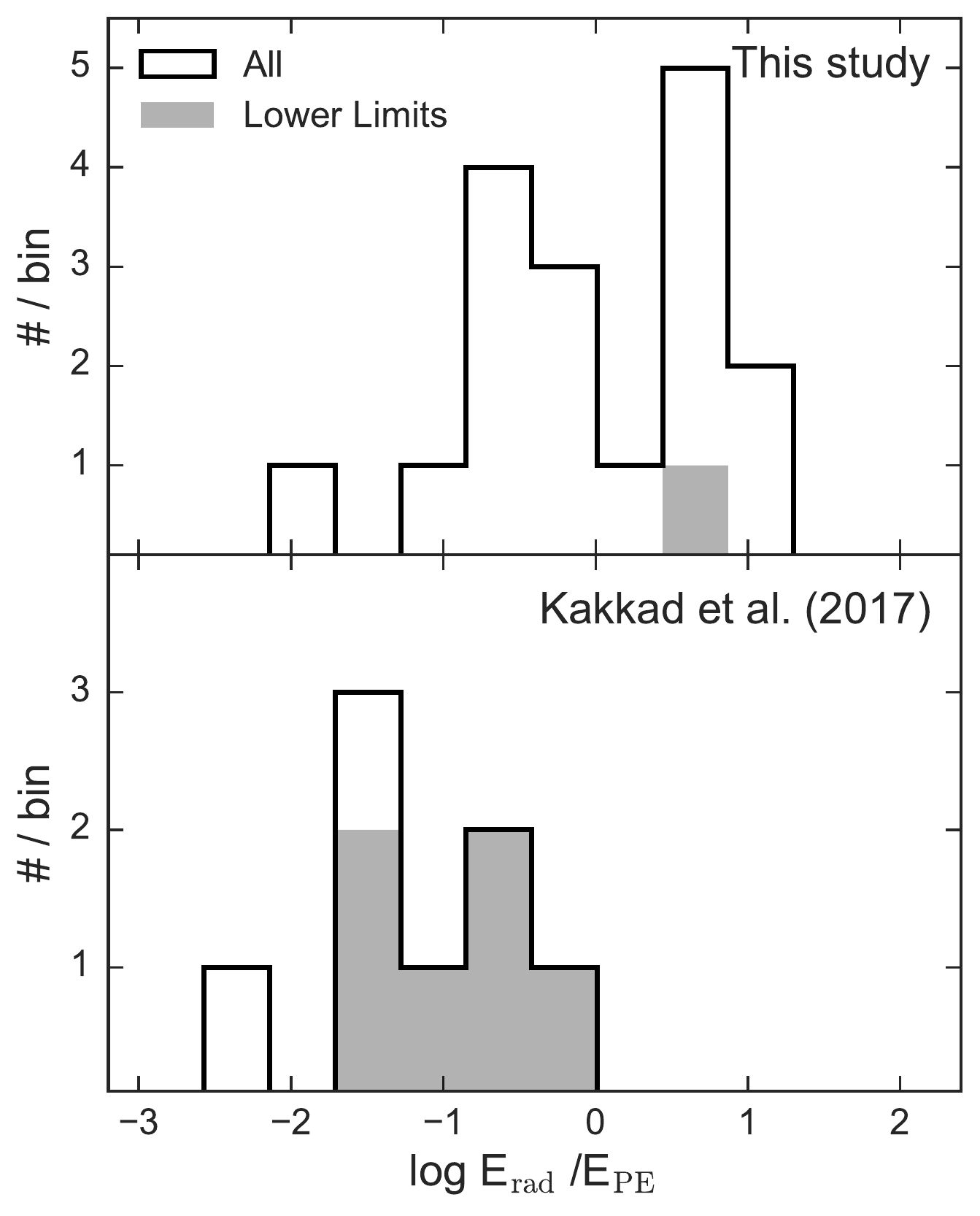}
\caption{The ratio of the radiative energy from the AGN expected to couple to the ISM ($E_{\rm rad}$) to the gravitational
potential energy of the observed molecular gas ($E_{\rm PE}$) among the LLAMA AGN (top panel) and the high-redshift AGN sample
from \citet{kakkad17} (bottom panel). Shaded histograms additionally show the subset of objects with lower limits to the ratio, because
they either have upper limits to the CO line or unconstrained stellar masses (Equation \ref{pote}). 
One object lies outside the plot to the right in the upper panel, and one lies outside to the left in the lower panel.
Among the LLAMA AGN, the ratio scatters about unity, implying that the radiative output from the AGN phase is 
sufficient to unbind or strongly disturb the molecular gas. In contrast, the ratio for the high-redshift AGN mostly lie well-below
unity, though most are lower limits.
}
\label{stability_calc}
\end{figure}

In the upper panel of Figure \ref{stability_calc}, we plot the ratio E$_{\rm rad}$/$\vert$E$_{\rm PE}\vert$ for the AGN. In cases where we only have
upper limits on \coline\ (MCG -05-23-016) or on the central stellar mass (NGC 3783), we use these limits to estimate a lower limit to 
E$_{\rm rad}$/$\vert$E$_{\rm PE}\vert$ (grey filled histogram).
While the ratio of energies shows considerable spread, there is sufficient radiative energy
in most of the AGN to dynamically influence the molecular gas, given our assumptions. This highlights the relevance of both the luminosities
of our AGN sample and the scales probed by our study. While these AGN may not be luminous enough to greatly
affect the galaxy-wide gas content of their hosts, they do have the capacity to destabilise the cold gas within their central
few kpc. When molecular material is more centrally concentrated than the gravitational potential of the galaxy, $\eta > 1$ and the gas is
easily destablised. This is the likely explanation for the powerful molecular outflows seen among local ULIRGs harbouring AGN 
\citep[e.g.,][]{feruglio10, cicone14}.  

\subsection{The central molecular gas in AGN host galaxies}

We have compared and contrasted the central \coline\ luminosities (\lco; Section \ref{co_lums}) and molecular 
gas fractions (f$_{\rm gas}$; Section \ref{gas_fracs}) of the LLAMA AGN and inactive control galaxies. Despite
the large spread of these quantities, we find them to be statistically indistinguishable between the subsamples.
The most noticeable difference is the larger number of CO non-detections among the
inactive group. Gas-poor galaxies may be under-represented among our AGN hosts, though, in both subsets, they
constitute a minority, and hence the difference in CO detection rates among AGN and inactive galaxies is still consistent
with sampling effects. We conclude that, for the most part, the galaxies in our sample have equivalent amounts of
cold gas in their centres irrespective of their level of nuclear activity.

The star formation efficiencies and cold gas depletion timescales (\taud) in the centres of LLAMA galaxies are comparable to those found in the disks
of normal inactive star forming galaxies in the local Universe (Section \ref{sf_efficiency}), as long as we assume a depressed \alphaco\ in these central regions
\citep{sandstrom13}. Broadly speaking, the AGN and inactive galaxies have very similar distributions of central \taud, and our
controlled study shows at best a marginal decrease in \taud\ (or an increase in SFE) in the AGN. This is reflected in the slight offset ($\times 3$)
between AGN and inactive galaxies in Figure \ref{lco_lgal_cen}. This offset is independent of the various uncertain factors needed to 
convert measured luminosities to gas masses or SFRs.

The clearest conclusion from our analysis is that AGN definitely do not show strongly deviant SFEs, even in gas that is within
the dynamical sphere of influence of a moderately luminous nuclear source. In particular, there is no evidence that the SFEs 
of the central molecular gas in AGN hosts are systematically suppressed. 
This implies that radiative feedback or winds, amply evident in the kinematics of emission lines from ionised gas and 
excited warm molecular gas in Seyferts \citep{veilleux05, mullersanchez11, bae14},
do not substantially alter the environments within molecular clouds that produce the bulk
of the low-excitation CO emission. We have shown above that there is ample energy in these typical Seyferts to influence
the dynamics of the gas. The normal values of the SFE in the LLAMA AGN suggests that the coupling of 
mechanical feedback energy from the nucleus with the cold, dense, star-forming phase in not as efficient
as canonical models have assumed, at least in the settled disk galaxies that host most Seyfert nuclei in the local Universe.

The geometry of the AGN radiation field and winds may be an important mitigating factor. CO-emitting gas lies primarily
in the plane of a galaxy's thin disk. We have assumed in the calculations of Section \ref{feedback_calcs} that this gas intersects with a substantial
portion of the absorbed bolometric power from the AGN. However,  most of the power may instead be 
absorbed and carried away in a hot phase that interacts only weakly with cold molecular clouds. This could
happen due to disk plane pressure gradients, which force hot winds to escape perpendicular 
to the disk as suggested by some high-resolution hydrodynamic simulations \citep{gabor14,wada16}.
The primary mode of AGN feedback on galaxies would then be
through the heating of galaxy atmospheres that strangulates future star formation, rather than through 
the prompt shut-down of existing star formation. 
 
While all of the CO-detected AGN have SFEs at or above the norm for star-forming disks, 
4 of the 14 CO-detected inactive galaxies have SFEs depressed by at least a factor of 5
compared to this norm. This may point to a situation where galaxies with low central SFEs do not easily 
host luminous AGN. While the result needs to be verified with more statistical rigour using larger samples,
we speculate on its implication in light of AGN fuelling mechanisms.
The SFE of bulk molecular gas is directly proportional to the fraction of the gas that can collapse into dense 
pre-stellar cores and sustain star formation. The processes that hinder the formation of dense cores in nuclear molecular
gas, such as high levels of turbulence, low molecular gas filling factors, or low gas pressures \citep{krumholz05}, also prevent this gas from effectively
forming an accretion flow to the central SMBH. Thus, a marginally unstable circum-nuclear molecular structure
may be a necessary prerequisite for AGN activity at a high accretion rate. 

\subsection{Comparisons to AGN host galaxies at high redshift}

Star formation during the redshift interval $1<z<3$
has particular relevance for our understanding of the evolution of galaxies, since most of the stellar mass
of the Universe was produced at that time \citep{madau14}. AGN are more frequent, and possibly more luminous, at
this epoch \citep[e.g.,][]{aird10}, so there is much interest in unraveling the signatures of AGN feedback on molecular gas in these distant
AGN hosts \citep[e.g.,][]{brusa15, kakkad17}.

Recently, \citet{kakkad17} reported lower \taud\ and molecular gas fractions from an ALMA survey of ten X-ray selected AGN at $z\sim1.5$. 
In the scenario of suppressive AGN feedback on star formation, 
their results are interpreted as widespread evidence for the prompt removal of star-forming molecular gas, possibly by an outflow,
which depresses $f_{\rm gas}$. A lower \taud\ results from this because the SFR tracer (in their case, the FIR continuum) responds
to the associated decrease in star formation more slowly that the timescale over which the gas is removed.

At face value, these results are inconsistent with our findings, since we find the same distribution of gas fractions in the LLAMA galaxies
independent of their state of nuclear activity. Therefore, we compare some of the relevant properties of the AGN studied in \citet{kakkad17}
to the AGN in LLAMA to help appreciate these differences.

In the lower panel of Figure \ref{stability_calc}, we plot the ratio E$_{\rm rad}$/$\vert$E$_{\rm PE}\vert$ ratio for the \citet{kakkad17} sample, calculated
using Equations \ref{erad} \& \ref{pote}. We use the $L_{bol}$, \smass, and $M_{H2}$ as tabulated in their work, and adopt
an R$_{\rm beam}$ of 4 kpc, consistent with the 1 arcsec ALMA resolution of their \coline\ maps. We have assumed that the gravitational
potential in these distant AGN is dominated by baryons (stars and gas). Seven of their targets only have upper limits in $M_{H2}$,
which we show as shaded histograms in Figure \ref{stability_calc}.

Assuming $\eta=1$ as in Section \ref{feedback_calcs}, 
we find that the AGN from \citet{kakkad17} generally have E$_{\rm rad}$/$\vert$E$_{\rm PE}\vert < 1$, and are therefore
less capable of disturbing their molecular gas than the LLAMA AGN.
This is because the host galaxies of these distant AGN are orders of magnitude more
gas-rich than the centres of local Seyfert galaxies. Even though the luminous output of the AGN from \citet{kakkad17}
are a factor of $\sim 10$ higher than those in LLAMA, they cannot produce enough energy to easily unbind 
the sheer amount of gas found in their massive, strongly star-forming hosts. This makes the strong 
evidence for AGN feedback reported in \citet{kakkad17} even more at odds with our findings.

There may be many reasons for this difference. The molecular gas in distant AGN hosts may be more centrally concentrated
than the stellar distribution ($\eta \gg 1$), though the \coline\ map is resolved in one of the targets from \citet{kakkad17}, suggesting
this is not universally the case. The coupling between the AGN luminosity and molecular gas ($\epsilon_{r}$) may be significantly
higher than canonical models imply. Alternatively, the timescale for luminous AGN activity (t$_{\rm AGN}$) may be
longer at these redshifts. Finally, more luminous AGN may be able to excite and destroy the CO molecule more effectively, which
will increase \alphaco, though a comparison of CO-based and dust-based gas masses for one well-studied case suggests
otherwise \citep{brusa15}. 
All these unknowns may play a part in enhancing the ability of nuclear activity to unbind galaxy-wide
gas and reconcile our results with those of \citet{kakkad17}. If so, this would imply a clear difference between local and high-redshift galaxies,
with respect to the detailed physics that governs the cross-section of interaction between the output of the AGN and 
the star-forming gas. It is essential to tie this down if the global importance of AGN feedback is to be fully understood.

\subsection{Is AGN obscuration related to the amount of cold gas in a galaxy's centre?} \label{obsc_trends}

We find a mild positive correlation between the nuclear obscuring column (N$_{H}$) and \icor\ (Section \ref{agn_corr}).
Our result is more significant than previous work from \citet{strong04}, though they used a set of AGN that
covered a much wider range in distance (15-300 Mpc) and their observations were not corrected for beam-dilution effects. 
The existence of this correlation suggests that the material that obscures at least some of the high energy 
emission from the AGN is linked to molecular gas in the
larger environment of the nucleus. In the popular AGN unification scheme, most nuclear obscuration can be attributed to
a pc-scale ``torus" with an anisotropic gas distribution (see \citet{netzer15} for a modern review). A small fraction
of the obscuration could come from gas on kpc-scales, as some IR studies have postulated \citep{deo07, goulding12, prieto14}.
A weak connection may also arise due to the long-lived gaseous inflow patterns between the torus and circum-nuclear gas, or due to
the molecular emission from the torus itself \citep{garciaburillo16, gallimore16}. Modern interferometric observatories such as ALMA
will obtain molecular line maps that resolve the inner circum-nuclear region ($\sim 100$ pc) for complete samples of nearby AGN. This
should allow a detailed assessment of whether the immediate nuclear environment and the covering factor of central molecular gas
plays a part in the obscuration properties of AGN.

The observed fraction of obscured AGN is a function of nuclear luminosity \citep[e.g.,][]{lawrence82,davies15}.
In our study, both obscured and unobscured AGN are equally luminous (Figure \ref{lco_lx}), but larger samples, typically 
spanning a wider range of distances, will suffer from unavoidable Malmquist bias. 
Such luminosity-dependent effects can masquerade as trends with obscuration, likely a limitation
of much earlier work. It is essential that future investigation into the connections between large-scale molecular gas and AGN obscuration is crafted
with special care to minimise such biases or account for them through modelling.

\subsection{Star-formation in AGN host galaxies}

AGN and inactive galaxies span a range of IR luminosities. AGN betray themselves in the form of clear MIR excesses, and we have 
used this fact to confidently separate the AGN and host galaxy IR contributions from the IR SEDs (Appendix \ref{agn_templates_discussion}). 
We find that AGN host galaxies are systematically more luminous in the FIR, which can be attributed to higher levels of star-formation even when controlling
for stellar mass and Hubble type. The equivalent galaxy-integrated SFRs of AGN hosts are $\approx 3\times$ higher than controlled
inactive galaxies, in a median sense. While we cannot take our results as independent evidence for higher SFRs in AGN due to the limited
statistical power of our sample, they are in line with a number of other studies which demonstrate that AGN, especially luminous ones, 
are preferentially found among star-forming and FIR-bright host galaxies \citep[e.g.,][]{salim07, rosario13, shimizu17}.

This observation highlights the important role that cold gas plays in the fueling of Seyfert AGN activity. The same gaseous reservoirs sustain
star formation in these galaxies. Even among galaxies with similar gross optical properties and structure, it is the ones that have enough
gas to sustain star formation that are also preferentially the hosts of AGN. This connection has been uncovered in X-ray luminous AGN hosts
across a broad range of redshifts \citep[e.g.,][]{santini12, rosario13, vito14}.

We have used the {\it WISE} 22 \mics\ images to determine the scaling to central IR luminosities for most of the inactive galaxies. Unlike
in the canonical FIR ($> 60$ \mics), the MIR dust emission could have a substantial component that arises from heating by the interstellar
radiation field from older stars \citep[e.g.][]{groves12}. This additional contribution boosts the central MIR emission relative to the extended MIR
emission and flattens the dust emission profiles of these galaxies in the MIR compared to the FIR. Our examination of the beam-matching
factor ($R_{ir}$) in Figure \ref{beam_correction} suggests that this effect is not large, since the distribution of $R_{ir}$ is very similar
between AGN and inactive galaxies, even though the factor was calculated from 160 \mics\ images for the AGN. However, even if there
is a small contribution to the 22 \mics\ emission from dust that is not heated by star formation, this will only heighten the difference in
the central SFRs between AGN and their control.

\section{Conclusions}

We have used the APEX and JCMT telescopes to obtain \coline\ spectroscopy 
for a matched sample of 17 AGN and 18 control galaxies from the LLAMA survey. Using this data, 
along with an additional measurement in the literature, we have explored whether the properties of the cold 
molecular gas in the central few kpc of AGN differ significantly from inactive galaxies with similar host characteristics.

{\bf 1)} The central \coline\ intensities (\ico),  gas fractions ($f_{\rm gas}$), and molecular gas depletion times (\taud)
of AGN are statistically indistinguishable from those of the inactive control galaxies.

{\bf 2)} The centres of AGN hosts and inactive control galaxies both lie on the same relationship between \lco\ and SFR
as the extended discs of star-forming galaxies, implying normal efficiencies of star formation in the central molecular gas.

{\bf 3)} These results indicate that AGN do not strongly influence the cold, star-forming molecular
phase in the centres of their hosts, despite heuristic arguments which suggest that the AGN in LLAMA
are luminous enough to dynamically disturb this material (Section \ref{feedback_calcs}). We conclude
that the prompt feedback from nuclear activity on the star-forming molecular phase is weak.

{\bf 4)} The most CO-rich AGN are also optically and X-ray obscured. This suggests a weak but definite
connection between kpc-scale cold gas and the obscuration to the nucleus, which may be {\it in situ} or mediated
by the dynamical connection between circum-nuclear material and the obscuring torus.
A larger survey of AGN is needed to explore this more effectively, and Section \ref{obsc_trends} discusses a possible strategy.

\section*{Acknowledgements}

We thank the anonymous referee for insight that has substantially improved the quality of this work.
D.R. and D.M.A. acknowledge the support of the UK Science and Technology Facilities Council (STFC) through grant ST/L00075X/1.
L.B. acknowledges support by a DFG grant within the SPP 1573 ``Physics of the interstellar medium".
M.K. acknowledges support from the Swiss National Science Foundation and Ambizione fellowship grant PZ00P2\textunderscore154799/1.
A.S. acknowledges the support of the Royal Society through the award of a University Research Fellowship.
The Atacama Pathfinder Experiment (APEX) is a collaboration between the Max-Planck-Institut f\"ur Radioastronomie, the 
European Southern Observatory, and the Onsala Space Observatory. The James Clerk Maxwell Telescope (JCMT) is operated by the 
East Asian Observatory on behalf of NAOJ, ASIAA, KASI, NAOC,
 and the Chinese Academy of Sciences (Grant No. XDB09000000), with additional funding support from the 
STFC and participating universities in the United Kingdom and Canada.
The Wide-field Infrared Survey Explorer (WISE) is a joint project of the University of California, Los Angeles, and 
JPL/California Institute of Technology, funded by NASA. The Infrared Astronomical Satellite (IRAS) was a joint project of the US, UK and the Netherlands.
Herschel is an ESA space observatory with science instruments provided by European-led Principal Investigator consortia and with important participation from NASA.

\bibliographystyle{mn2e}

\bibliography{mn-jour,agnlp_co}

\appendix

\section{Empirical AGN SED templates in the Infrared} \label{agn_templates_discussion}

\begin{figure}
\includegraphics[width=\columnwidth]{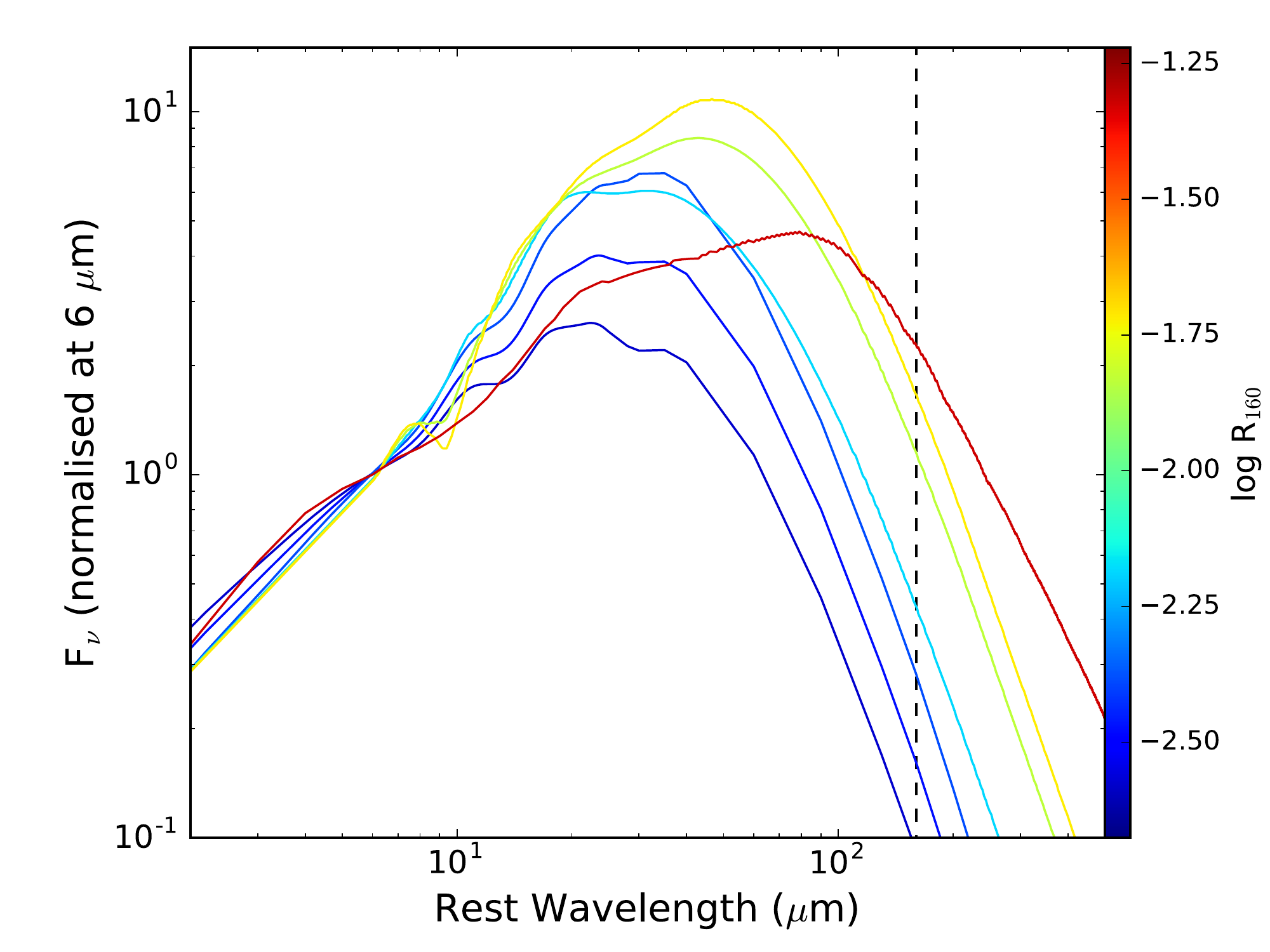}
\caption{The library of empirical IR SED templates of AGN used in this work. The SEDs are coloured by their single shape
parameter $R_{160}$, the ratio of the rest-frame 160 \mics\ luminosity to the integrated 8--1000 \mics\ luminosity (color
key on the right). 
}
\label{agn_templates}
\end{figure}

A number of recent studies have used MIR spectroscopy from the Spitzer/Infra-Red Spectrograph (IRS) and 
FIR photometry from {\it IRAS} to arrive at the intrinsic AGN-heated dust
emission in representative samples of local Seyfert galaxies and QSOs. Despite a similarity in approach, there remains 
some disagreement among these studies on the degree to which the light from the AGN is reprocessed by cold dust that lies
 $\gtrsim$ kpc from the nucleus. While most empirical AGN SED templates agree on the general shape of
the AGN torus emission at their peak wavelengths ($\lesssim 20$ \mics), they differ by as much as an order of magnitude
in the long-wavelength tail emitted by AGN-heated dust in the FIR. In general, the templates derived by \citet{mor12} imply a steep drop-off
to long wavelengths, while the template recently proposed by \cite{symeonidis16}, applicable to luminous PG QSOs, indicates a much
slower drop-off, with more of the AGN's IR luminosity arising in the FIR. The popular templates from \citet{mullaney11}, based
on fits to {\it SWIFT}-BAT AGN,  exhibit a behaviour intermediate to the templates of \citet{mor12} and \citet{symeonidis16} (Figure \ref{agn_templates}).

Rather than restrict ourselves to a choice of AGN templates that is dictated by one study, 
we instead considered the ensemble of all templates compiled by the works listed above --
three each from 
\citet{mor12}\footnote{The upper and lower scatter about the mean \citet{mor12} template was kindly provided by H. Netzer (private communication).}
and \citet{mullaney11}\footnote{The \citet{mullaney11} templates were extrapolated to 1 \mics\ with a power-law of fixed slope that matches those of the other template families (Figure \ref{agn_templates}).} representing the median and scatter shown by the AGN in their samples, and the mean SED from \citet{symeonidis16}. These seven templates can be cast into a sequence delineated by the fraction of the IR luminosity that is emitted in the FIR. 
Therefore, we define a parameter $R_{160} \equiv L_{160} /$\lagn, where $L_{160}$ is the monochromatic rest-frame
160 \mics\ luminosity of the AGN template, and \lagn\ is its integrated 8--1000 \mics\ luminosity. We plot the final set of seven templates
in Figure \ref{agn_templates} differentiated by $R_{160}$. This library of basis templates was used in the SED fits described in Section \ref{sed_fits}.

\begin{figure*}
\includegraphics[width=\textwidth]{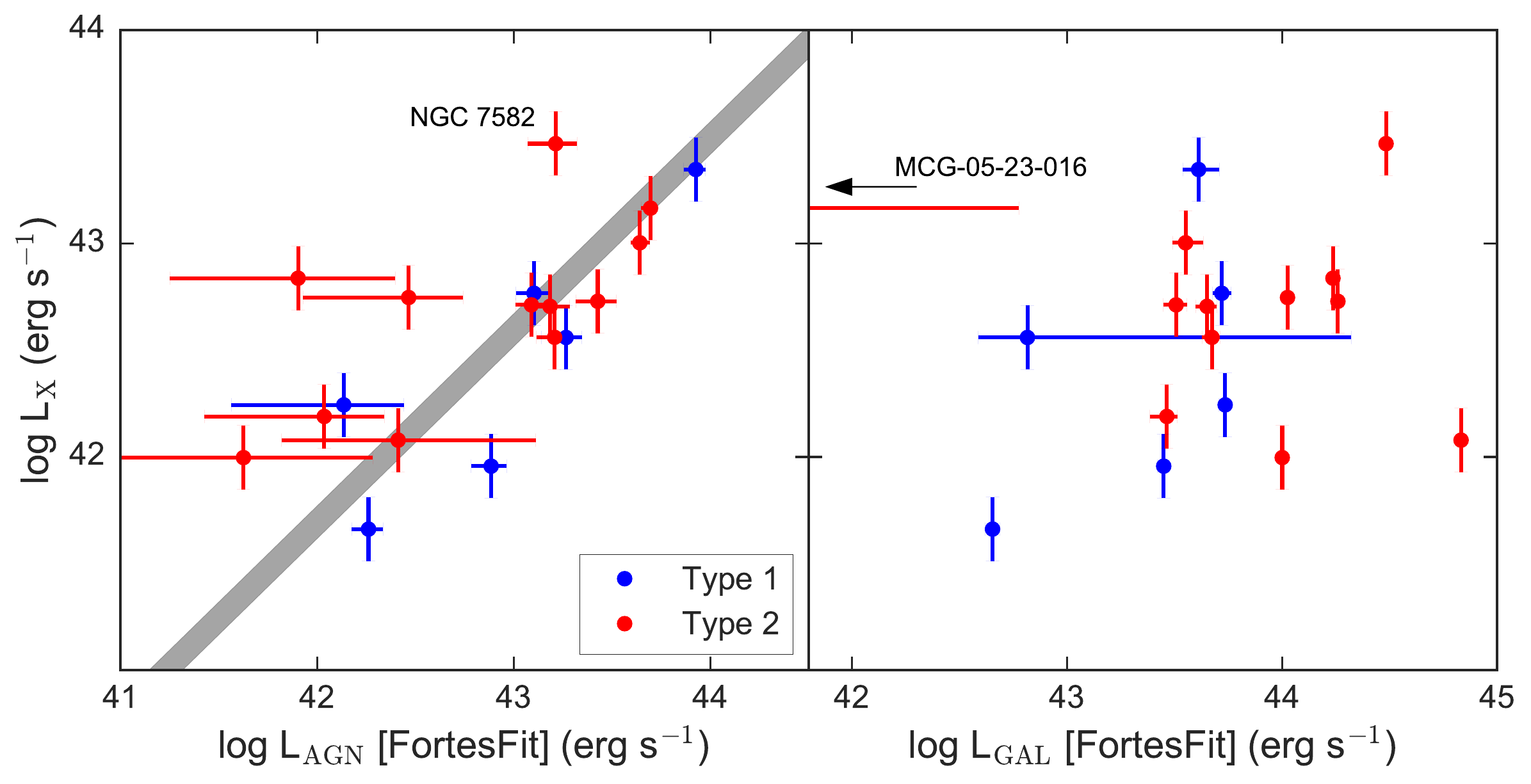}
\caption{{\bf Left:} The intrinsic 2-10 keV X-ray luminosity (\lx) of the LLAMA AGN plotted against the 8--1000 \mics\ AGN luminosity (\lagn)
derived from SED fits. The grey shaded band shows the expected trend based on the best-fit X-ray -- MIR relationship from \citet{gandhi09}
over all AGN IR templates used in the fits (Section \ref{sed_fits}). NGC 7582 is the largest outlier, but it is known to show strong variability
in its X-ray absorbing column. Type 1(2) AGN are differentiated by blue (red) points.
{\bf Right:} The intrinsic 2-10 keV X-ray luminosity (\lx) of the LLAMA AGN plotted against the 8--1000 \mics\ luminosity of the galaxy (\lgal)
derived from SED fits. 
}
\label{lx_lagn_lgal}
\end{figure*}

In Section \ref{ir_luminosities}, we explore the differences between the galaxy IR luminosities (\lgal) between AGN and control galaxies.
Errors in this analysis could arise if the AGN IR templates do not adequately capture the emission from cold dust that is heated 
directly by the AGN. As discussed above, the inclusion of a range of empirical AGN IR template shapes is motivated by
our current best understanding of the diversity of the long-wavelength emission from AGN. However, one may
argue that this approach is still not wholly satisfactory. The AGN that were used to construct the empirical AGN templates
are typically more luminous than the LLAMA AGN, and, regardless of the method, the FIR emission of the templates
are still extrapolations, since a model dust component from the galaxy dominated the FIR in all these studies. Therefore, the possibility remains
that the luminosity of AGN-heated dust in the LLAMA AGN was underestimated, and a portion of the true emission
from the AGN was incorrectly incorporated into \lgal\ by the SED fits. If this were the case, we 
would expect a correlation between \lgal\ and \lx, an independent measure of the luminosity of the AGN.

Figure \ref{lx_lagn_lgal} tests this notion. In the left panel, we demonstrate that the IR luminosity of the AGN (\lagn) 
derived from the SED fits correlates with \lx, and that the slope of the trend is consistent with the expectation from
the X-ray--MIR relation of \citet{gandhi09}. A few poorly-constrained estimates of \lagn\ tend to be significantly lower
than the expectation from this relation (grey band), but an inspection of their fits show that, in all cases, the AGN 
is faint with respect to the MIR emission from the system and does not display much contrast in the combined SED.
The only object in which an energetically important AGN contribution may be underestimated is NGC 7582 (marked in the diagram).
This well-known Seyfert 2 is known to display strong short-timescale X-ray spectral variations from a complex set of absorbers
\citep{piconcelli07}, and recent NuStar analysis \citep{rivers15} suggest an intrinsic flux that is an order of magnitude lower than the
value we used for this study. Taking all this information together, we consider the general agreement between our estimates of 
\lagn\ and the predictions from the \lx\ as evidence for the validity of our fits.

Unlike with \lagn, there is no correlation between \lx\ and \lgal\ (right panel of Figure \ref{lx_lagn_lgal}). In order to resolve the
differences in the distributions of \lgal\ between AGN and inactive galaxies in Figure \ref{lir_integ_dists}, 
most of the AGN would have required a boost of $\gtrsim3\times$ to their FIR luminosities, 
which would betray itself clearly in this figure in the form of higher \lgal\ among the more
luminous AGN. The lack of any significant correlation implies that we have not systematically included -- or excluded -- 
any important portion of an AGN's IR luminosity incorrectly as emission from the galaxy.

\section{MCG -05-23-016}

\begin{figure}
\includegraphics[width=\columnwidth]{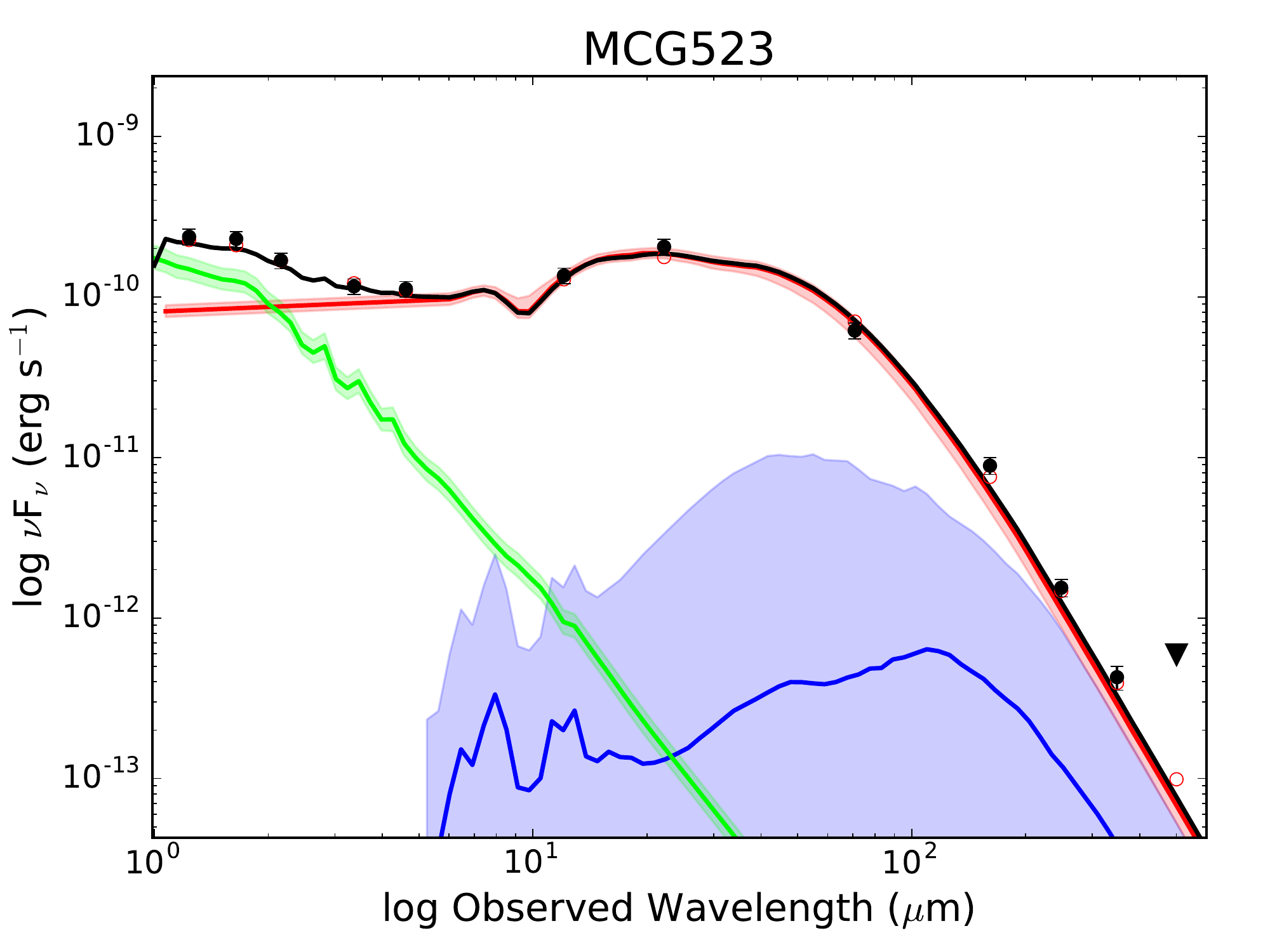}
\caption{The IR SED of MCG -05-23-016. Black points mark the photometry, with detections (filled circles) and an upper limit (downward arrowhead).
The results of the multi-component SED fit for MCG -05-23-016 are also shown, consisting of a stellar component (green), an AGN component (red),
and a galaxy dust-emission component (blue) (see Section \ref{sed_fits} for more details about these components
and the fitting methodology). Shaded regions show the 16th and 84th percentile scatter about the median model SED (solid coloured lines).
The black solid line is the sum of all the best-fit components. The photometry of the best-fit model is shown with red open circle points.
The IR SED to sub-mm wavelengths can be reproduced with a pure AGN component. The galaxy dust component is purely an upper bound.
}
\label{mcg523_fortesfit}
\end{figure}

This relatively powerful AGN (\lx$=2\times10^{43}$ \ergs) has been studied extensively in recent years due to its
complex X-ray spectrum, which shows a strong Fe K$\alpha$ fluorescence line from both the accretion disk and torus,
as well as a prominent reflection hump indicating substantial amounts of nuclear cold gas \citep[e.g.][]{balokovic15}.
It is hosted by a gas-poor early-type host galaxy with a low stellar mass \smass$\approx 2\times10^{9}$ \msun.
This AGN is unique in our sample because its SED is consistent with pure AGN-heated dust even into the sub-mm bands,
with no hint of a cold dust excess from star-formation (Figure \ref{mcg523_fortesfit}). In this sense, it provides us with one of the
best constraints on the intrinsic torus emission from an AGN in the nearby Universe. We find a best-fitting
value of $R_{160} = 1.96^{+0.33}_{-0.88}\times10^{-2}$ which is substantially shallower towards the FIR than the mean empirical AGN SED
of \citet{mor12}, but consistent with the flatter SEDs presented by \citet{mullaney11}. It is also somewhat steeper
than the FIR-bright SED proposed by \citet{symeonidis16}. In this galaxy, we can assert with some confidence that
the AGN is not responsible for heating a great deal of extended cold dust which could be mistakenly attributed to star formation.
Having said this, MCG -05-23-016 is the only AGN in LLAMA that is not detected in CO. The $2\sigma$ upper limit on its central
gas mass is $M_{H2} > 3.6\times10^{7}$ \msun, sufficient to fuel the AGN at its current luminosity for 700 Myr. Therefore,
the bright and prominent torus is still quite consistent with our constraints on the molecular emission from this system.
MCG -05-23-016 has an unusually small and faint narrow line region \citep{prieto14}. Therefore, it is  possible that the paucity of gas 
and dust in the central kpc is responsible for the lack of extended AGN heated dust emission. Much of the detectable gas in this galaxy
appears to be confined to within the inner few 100 pc.

\end{document}

%% file: Table1.tex
\begin{table*}
\caption{Basic data and $^{12}$CO 2--1 flux measurements for the LLAMA galaxies in this work}
\label{basic_data}
\begin{tabular}{cccccc}
\hline \hline
Name & Distance & AGN Type & Telescope & I$_{\rm CO}$ [2$\rightarrow$1]$^{\rm a}$ & S$_{\rm CO}$ \\
 & (Mpc) &  &  & (K km s$^{-1}$) & (Jy km s$^{-1}$) \\
 \hline 
 \multicolumn{6}{c}{AGN} \\
 \hline
ESO 021-G004 & 39 & 2 & APEX & $    4.5\pm1.0 $ & $   132\pm15 $ \\
ESO 137-G034 & 35 & 2 & APEX & $    3.0\pm0.0 $ & $    89\pm10 $ \\
MCG-05-23-016 & 35 & 1i & APEX & $< 1.3    $ & $< 38    $ \\
MCG-06-30-015 & 27 & 1.2 & APEX & $    1.0\pm0.0 $ & $    29\pm5  $ \\
NGC 1365 & 18 & 1.8 & SEST & $  150.0\pm10.0$ & $  3075\pm205$ \\
NGC 2110 & 34 & 1i & JCMT & $    3.4\pm0.0 $ & $    57\pm6  $ \\
NGC 2992 & 36 & 1i & JCMT & $   22.4\pm2.0 $ & $   377\pm28 $ \\
NGC 3081 & 34 & 1h & JCMT & $    4.8\pm1.0 $ & $    80\pm17 $ \\
NGC 3783 & 38 & 1.5 & APEX & $    3.5\pm0.0 $ & $   103\pm12 $ \\
NGC 4235 & 37 & 1.2 & APEX & $    2.3\pm0.0 $ & $    68\pm9  $ \\
NGC 4388 & 25 & 1h & JCMT & $   22.3\pm2.0 $ & $   374\pm29 $ \\
NGC 4593 & 37 & 1.0 & JCMT & $   10.0\pm2.0 $ & $   168\pm28 $ \\
NGC 5506 & 27 & 1i & JCMT & $   10.1\pm1.0 $ & $   169\pm20 $ \\
NGC 5728 & 39 & 2 & JCMT & $   21.9\pm2.0 $ & $   368\pm41 $ \\
NGC 6814 & 23 & 1.5 & JCMT & $    5.7\pm1.0 $ & $    96\pm22 $ \\
NGC 7172 & 37 & 1i & APEX & $   18.7\pm1.0 $ & $   548\pm28 $ \\
NGC 7213 & 25 & 1 & APEX & $    8.2\pm1.0 $ & $   240\pm16 $ \\
NGC 7582 & 22 & 1i & APEX & $   95.8\pm3.0 $ & $  2803\pm83 $ \\
\hline 
\multicolumn{6}{c}{Inactive Galaxies} \\
\hline
ESO 093-G003 & 22 &  & APEX & $   19.9\pm1.0 $ & $   581\pm38 $ \\
ESO 208-G021 & 17 &  & APEX & $< 0.4    $ & $< 12    $ \\
IC 4653 & 26 &  & APEX & $    3.6\pm0.0 $ & $   107\pm9  $ \\
NGC 1079 & 19 &  & APEX & $    0.6\pm0.0 $ & $    19\pm5  $ \\
NGC 1947 & 19 &  & APEX & $   12.0\pm1.0 $ & $   351\pm26 $ \\
NGC 2775 & 21 &  & APEX & $< 0.4    $ & $< 12    $ \\
NGC 3175 & 14 &  & APEX & $   31.4\pm1.0 $ & $   918\pm23 $ \\
NGC 3351 & 11 &  & APEX & $   37.7\pm1.0 $ & $  1104\pm22 $ \\
NGC 3717 & 24 &  & APEX & $   30.1\pm1.0 $ & $   881\pm29 $ \\
NGC 3749 & 42 &  & APEX & $   15.7\pm1.0 $ & $   459\pm34 $ \\
NGC 4224 & 41 &  & APEX & $    4.1\pm0.0 $ & $   120\pm11 $ \\
NGC 4254 & 15 &  & APEX & $   34.1\pm1.0 $ & $   998\pm30 $ \\
NGC 4260 & 31 &  & APEX & $< 1.1    $ & $< 31    $ \\
NGC 5037 & 35 &  & APEX & $    9.8\pm1.0 $ & $   286\pm27 $ \\
NGC 5845 & 25 &  & APEX & $< 0.9    $ & $< 26    $ \\
NGC 5921 & 21 &  & APEX & $   11.6\pm0.0 $ & $   340\pm14 $ \\
NGC 718 & 23 &  & APEX & $    1.9\pm0.0 $ & $    57\pm4  $ \\
NGC 7727 & 26 &  & APEX & $    2.3\pm0.0 $ & $    68\pm8  $ \\
\hline
\multicolumn{6}{l}{\textsuperscript{a}\footnotesize{Scaled to the T$_{mb}$ scale.}} \\
\end{tabular}
\end{table*}

%% file: Table2.tex
\begin{table*}
\caption{Various derived quantities for the LLAMA galaxies in this work}
\label{derived_data}
\begin{tabular}{ccccccc}
\hline \hline
Name & $\log$ L$^{\prime}_{\rm CO}$ & $\log$ L$_{\tiny GAL}$ & $\log$ L$_{\tiny AGN}$ & $\log$ L$_{\rm X}$ & $\log$ N$_{\rm H}$ & $\log$ L$_{\rm K,AGN}$ \\
 & (K km s$^{-1}$ pc$^{2}$) & (erg s$^{-1}$) & (erg s$^{-1}$) & (erg s$^{-1}$) & (cm$^{-2}$) & (erg s$^{-1}$) \\
\hline 
\multicolumn{6}{c}{AGN} \\
\hline
ESO 021-G004 & $8.083\pm0.075$ & $43.45^{+0.07}_{-0.15}$ & $42.30^{+0.39}_{-0.32}$ & 42.19 & 23.8 & $<$ 42.10 \\
ESO 137-G034 & $7.820\pm0.076$ & $43.68^{+0.04}_{-0.05}$ & $43.23^{+0.08}_{-0.09}$ & 42.34 & 24.3 & $<$ 40.77 \\
MCG-05-23-016 & $< 7.445$ & $41.28^{+1.29}_{-1.29}$ & $43.70^{+0.03}_{-0.03}$ & 43.16 & 22.2 & 43.00 \\
MCG-06-30-015 & $7.109\pm0.090$ & $42.74^{+0.17}_{-0.17}$ & $43.27^{+0.04}_{-0.06}$ & 42.56 & 20.9 & 42.35 \\
NGC 1365 & $8.782\pm0.066$ & $44.82^{+0.02}_{-0.05}$ & $43.05^{+0.42}_{-1.03}$ & 42.31 & 22.2 & 42.63 \\
NGC 2110 & $7.603\pm0.073$ & $43.64^{+0.05}_{-0.13}$ & $43.20^{+0.19}_{-0.08}$ & 42.65 & 22.9 & 42.69 \\
NGC 2992 & $8.472\pm0.067$ & $43.99^{+0.03}_{-0.04}$ & $42.33^{+0.40}_{-0.84}$ & 42.11 & 21.7 & 42.20 \\
NGC 3081 & $7.749\pm0.105$ & $43.46^{+0.06}_{-0.07}$ & $43.25^{+0.08}_{-0.09}$ & 42.94 & 23.9 & 41.32 \\
NGC 3783 & $7.955\pm0.076$ & $43.57^{+0.10}_{-0.08}$ & $43.96^{+0.04}_{-0.05}$ & 43.23 & 20.5 & 43.39 \\
NGC 4235 & $7.754\pm0.080$ & $42.63^{+0.05}_{-0.06}$ & $42.36^{+0.09}_{-0.12}$ & 41.94 & 21.3 & $<$ 42.10 \\
NGC 4388 & $8.151\pm0.068$ & $44.25^{+0.03}_{-0.03}$ & $43.51^{+0.10}_{-0.09}$ & 43.20 & 23.5 & 41.75 \\
NGC 4593 & $8.146\pm0.090$ & $43.71^{+0.05}_{-0.07}$ & $43.20^{+0.11}_{-0.11}$ & 42.91 & $<$ 20.0 & 42.42 \\
NGC 5506 & $7.874\pm0.076$ & $43.62^{+0.09}_{-0.13}$ & $43.61^{+0.09}_{-0.11}$ & 43.10 & 22.4 & 43.09 \\
NGC 5728 & $8.531\pm0.075$ & $44.23^{+0.02}_{-0.02}$ & $42.22^{+0.39}_{-0.70}$ & 43.14 & 24.1 & $<$ 41.02 \\
NGC 6814 & $7.491\pm0.108$ & $43.73^{+0.02}_{-0.06}$ & $42.36^{+0.60}_{-0.41}$ & 42.32 & 21.0 & 41.73 \\
NGC 7172 & $8.658\pm0.063$ & $44.01^{+0.03}_{-0.03}$ & $42.74^{+0.21}_{-0.31}$ & 42.84 & 22.9 & 42.46 \\
NGC 7213 & $7.959\pm0.066$ & $43.43^{+0.03}_{-0.04}$ & $42.97^{+0.12}_{-0.10}$ & 42.06 & $<$ 20.0 & 42.46 \\
NGC 7582 & $8.917\pm0.061$ & $44.48^{+0.02}_{-0.03}$ & $43.29^{+0.10}_{-0.13}$ & 42.90 & 24.3 & 42.78 \\
\hline 
\multicolumn{6}{c}{Inactive Galaxies} \\
\hline
ESO 093-G003 & $8.233\pm0.065$ & $43.91^{+0.04}_{-0.04}$ & $< 40.8$ &  &  &  \\
ESO 208-G021 & $< 6.340$ & $41.93^{+0.13}_{-1.27}$ & $< 41.3$ &  &  &  \\
IC 4653 & $7.642\pm0.068$ & $43.15^{+0.04}_{-0.04}$ & $< 41.7$ &  &  &  \\
NGC 1079 & $6.610\pm0.118$ & $42.51^{+0.04}_{-0.05}$ & $< 39.6$ &  &  &  \\
NGC 1947 & $7.888\pm0.067$ & $42.72^{+0.04}_{-0.03}$ & $< 39.5$ &  &  &  \\
NGC 2775 & $< 6.520$ & $43.32^{+0.03}_{-0.04}$ & $< 40.6$ &  &  &  \\
NGC 3175 & $8.040\pm0.061$ & $43.68^{+0.03}_{-0.03}$ & $< 40.0$ &  &  &  \\
NGC 3351 & $7.911\pm0.061$ & $43.55^{+0.02}_{-0.02}$ & $< 39.4$ &  &  &  \\
NGC 3717 & $8.489\pm0.062$ & $43.96^{+0.02}_{-0.03}$ & $< 40.7$ &  &  &  \\
NGC 3749 & $8.691\pm0.068$ & $43.86^{+0.04}_{-0.03}$ & $< 40.7$ &  &  &  \\
NGC 4224 & $8.086\pm0.070$ & $42.93^{+0.06}_{-0.18}$ & $< 42.0$ &  &  &  \\
NGC 4254 & $8.134\pm0.061$ & $44.84^{+0.02}_{-0.02}$ & $< 40.5$ &  &  &  \\
NGC 4260 & $< 7.258$ & $42.35^{+0.07}_{-0.29}$ & $< 40.8$ &  &  &  \\
NGC 5037 & $8.328\pm0.071$ & $43.06^{+0.03}_{-0.04}$ & $< 40.1$ &  &  &  \\
NGC 5845 & $< 6.999$ & $41.69^{+0.17}_{-0.28}$ & $< 40.9$ &  &  &  \\
NGC 5921 & $7.960\pm0.062$ & $43.40^{+0.04}_{-0.04}$ & $< 40.7$ &  &  &  \\
NGC 718 & $7.262\pm0.067$ & $42.66^{+0.04}_{-0.07}$ & $< 38.8$ &  &  &  \\
NGC 7727 & $7.449\pm0.075$ & $42.56^{+0.07}_{-0.28}$ & $< 41.2$ &  &  &  \\
\end{tabular}
\end{table*}

%% file: Table3.tex
\begin{table}
\center
\caption{Summary of comparative statistics for AGN and Inactive galaxies}
\label{stats_table}
{\renewcommand{\arraystretch}{1.1}
\begin{tabular}{|c|c|c|c|}
\hline \hline
Quantity & Ensemble & Median$^{\rm a}$ & P$_{\rm test}$$^{\rm b}$  \\
 (Unit)    &  &  &  \\
\hline 
log I$_{\rm CO, 25}$ & AGN & $0.87^{+0.03}_{-0.03}$ \\
(log K km s$^{-1}$) & Inactive & $0.88^{+0.02}_{-0.02}$ & 0.43 \\
                                & Pairs & $0.35^{+0.15}_{-0.14}$ & 0.09 \\
\hline 
log f$_{\rm gas}$ & AGN & $-1.62^{+0.15}_{-0.14}$ & \\
                  & Inactive & $-1.65^{+0.17}_{-0.17}$ & 0.43 \\
                  & Pairs & $\phantom{-}0.30^{+0.28}_{-0.30}$ & 0.18 \\ 
\hline 
L$_{\rm GAL}$ & AGN & $43.64^{+0.06}_{-0.06}$ & \\
(log erg s$^{-1}$)  & Inactive & $43.21^{+0.05}_{-0.08}$ & 0.10 \\
                    & Pairs & $\phantom{-}0.60^{+0.20}_{-0.20}$ & 0.11 \\
\hline 
t$_{\rm dep}$ & AGN & $\phantom{-}9.10^{+0.20}_{-0.20}$ & \\
(log Years)   & Inactive & $\phantom{-}9.17^{+0.21}_{-0.21}$ & 0.43 \\
              & Pairs & $-0.15^{+0.37}_{-0.37}$ & 0.48 \\

\hline 
\multicolumn{4}{l}{\textsuperscript{a}\footnotesize{Median logarithmic difference for pairs.}}\\
\multicolumn{4}{l}{\textsuperscript{b}\footnotesize{Probabilities from 1) Kolmogorov-Smirnov test that AGN and}}\\
\multicolumn{4}{l}{\footnotesize{  control distributions are indistinct, and 2) Student T-test that}}\\
\multicolumn{4}{l}{\footnotesize{  differences of pairs are normally distributed with zero mean.}}\\
\end{tabular}}
\end{table}

%% file: Table4.tex
\begin{table*}
\caption{Basic data for Supplementary AGN}
\label{supplementary_agn}
\begin{tabular}{ccccccccl}
\hline \hline
Name & Distance & AGN Type & $\log$ L$_{\rm X}$ & $\log$ N$_{\rm H}$ & $r_{25}$ & inclination & I$_{\rm CO}$ [2$\rightarrow$1]$^{\rm a}$ & Reference for CO measurement \\
 & (Mpc) &  & (erg s$^{-1}$) & (cm$^{-2}$) & (arcsec) & (degree) & (K km s$^{-1}$) & \\
\hline 
NGC 1068 & 14.4 & 2.0 & 42.82 & 25.0 & 212 & 32 & $133.0\pm1.0$ & \citet{curran01} \\
NGC 3079 & 20.2 & 2.0 & 41.51 & 25.1 & 238 & 80 & $23.0\pm1.0$ & \citet{aalto95} \\
NGC 3227 & 23.0 & 1.5 & 42.37 & 21.0 & 161 & 47 & $31.7\pm1.8$ & \citet{rigopoulou97} \\
NGC 3516 & 38.9 & 1.2 & 42.75 & 20.0 &  52  & 39 & $<1.53$ & \citet{monje11} \\
NGC 4151 & 19.0 & 1.5 & 42.56 & 22.7 & 189 & 45 & $1.56\pm0.5$ & \citet{rigopoulou97} \\
NGC 5643 & 16.9 & 2.0 & 42.41 & 25.4 & 137 & 29 & $15.9\pm1.0$ & \citet{monje11}  \\
NGC 7314 & 17.4 & 1.9 & 42.18 & 21.6 & 137 & 63 & $2.8\pm0.4$ & \citet{rigopoulou97} \\
NGC 7465 & 27.2 & 1.9 & 41.93 & 21.5 &  37  & 50 & $3.1\pm0.6$ & \citet{monje11}  \\
\hline 
\multicolumn{9}{l}{\textsuperscript{a}\footnotesize{On the T$_{mb}$ scale.}} \\
\end{tabular}
\end{table*}